\documentclass{JHEP3}

\usepackage{amsfonts}
\usepackage{amssymb}
\usepackage[centertags]{amsmath}
\usepackage{amsthm}
\usepackage{ccaption}
\usepackage{mathrsfs}
\usepackage{graphicx}

\renewcommand{\vec}[1]{\textbf{#1}}
\newcommand{\AdS}[1][5]{\ensuremath{\mathrm{AdS}_{#1}} }
\newcommand{\RNAdS}[1][5]{\ensuremath{\mathrm{RN-AdS}_{#1}} }

\newcommand{\Sch}[1][5]{\ensuremath{\mathrm{Schr}_{#1}} }
\newcommand{\cSch}[1][5]{\ensuremath{\mathrm{RN-Schr}_{#1}} }
\newcommand{\Schr}{Schr\"{o}dinger~}
\newcommand{\AdSnrCFT}{holographic dictionary in \Schr space-times~}
\newcommand{\Schrcomma}{Schr\"{o}dinger,~}
\newcommand{\vol}[1]{\ensuremath{\mathrm{vol}_{#1}}}
\newcommand{\dimB}{\left(d+2\right)} 
\newcommand{\dimS}{\left(d+1\right)} 
\newcommand{\Prt}{\ensuremath{\mathrm{Pr}}}
                                     
\title{Charged, conformal non-relativistic hydrodynamics}
\author{Daniel K. Brattan \\
				Centre for Particle Theory \& Department of Mathematical Sciences, \\
				Science Laboratories, South Road, Durham DH1 3LE, United Kingdom. \\
			  Email: \email{d.k.brattan@durham.ac.uk} }
\preprint{}

\abstract{\ We embed a holographic model of an $U(1)$ charged fluid with Galilean invariance in string theory and calculate its specific heat capacity and Prandtl number. Such theories are generated by a $R$-symmetry twist along a null direction of a $\mathcal{N}=1$ superconformal theory. We study the hydrodynamic properties of such systems employing ideas from the fluid-gravity correspondence.}

\begin{document}

\section{Introduction}

{\ Systems at strong coupling continue to be one of the least tamed areas of modern physics. This is particularly puzzling considering the number of physical situations that seem to be described in terms of such models. Examples include non-BCS superconductors at the quantum critical point, the hydrodynamics of strongly coupled microscopic theories or the quark-gluon plasma (QGP). More curious still is the fact that until relatively recently we have even lacked a description of strongly coupled toy models, for instance, planar $\mathcal{N}=4$ SYM with a large 't Hooft coupling.}

{\ Notably, hydrodynamics has been the subject of intense study and, while there have been significant achievements in this area such as those of Navier, Stokes and Kolmogorov, there remain many phenomena, like turbulence, that still lack a full theoretical description. Importantly, an ability to completely describe the generic behaviour of fluids would lead to a deeper understanding of a vast range of models. This is because a typical feature of field theories with a long-wavelength expansion is a sector well described by the hydrodynamic regime.}

{\ For a fluid description whose microscopic origin is strongly coupled we cannot apply the usual perturbative methods to calculate approximate transport coefficients. As such, these fluids represent an interesting theoretical challenge. Moreover, beyond purely theoretical considerations, there are several practical applications where controllable models of such fluids would be useful. One example concerns the dynamics of the QGP whose transport coefficients are now accessible to experiment \cite{Adcox:2004mh, Back:2004je, Arsene:2004fa}. The interest in applying the gauge-gravity duality to this system comes from the fact that the correspondence describes a large class of ideal fluids and the QGP seems to be approximately ideal \cite{Schafer:2009dj}. Using the general lessons learned from the application of AdS/CFT has already led to a qualitative improvement in understanding the properties of this strange state of matter like its low viscosity to entropy ratio \cite{Nastase:2005rp}.}

{\ The AdS/CFT correspondence \cite{Maldacena:1997re,Gubser:1998bc,Witten:1998qj} provides a useful tool via which we can learn general lessons about certain strongly-coupled systems from their gravity duals. To date the most extensive use of this conjecture has been in the study of non-abelian gauge theories since early examples of the correspondence related $\mathcal{N}=4$ SYM, and similar models, to Einstein gravity. Like several dualities it relates a theory with a large expansion parameter to another theory with a small perturbation parameter. However, unusually for such a correspondence, the ``master field'' corresponding to the small curvature regime lives in one higher spacetime dimension than the strongly coupled theory, indicating a holographic description.}

{\ This paper is principally concerned with the fluid-gravity sector of the AdS/CFT correspondence where it was discovered that the restriction of certain large $N$ gauge theories to a long-wavelength regime is dual to a simplified gravitational description in which the relevant quantities are determined completely by hydrodynamic conservation laws \cite{Rangamani:2009xk}. This programme of studying duals to the fluid description of strongly coupled conformal field theories began with the seminal works of \cite{Policastro:2002se,Policastro:2002tn} where calculations of the graviton retarded Green's functions were made at the linearised level of gravity. A significant achievement in \cite{Bhattacharyya:2008jc} was a procedure to extend the previous calculations perturbatively to higher orders in derivatives of fluid velocity. It should be noted that the results obtained by these methods can give only qualitative predictions about nature because, as yet, no observed phenomenon is known to have as an underlying description a large $N$ gauge field.}

{\ Recently, significant effort has been expended in studying the non-relativistic limit of strongly-coupled field theories and their gravity duals. The interest in these theories in part lies in the fact that it has proven difficult to find generalisations of the AdS/CFT correspondence to other asymptotics. One of the few cases where this has been managed are the ``asymptotically \Schr spacetimes'' where the dual field theory satisfies the \Schr algebra \cite{Hagen:1972pd, Mehen:1999nd, Nishida:2007pj} possibly with a central extension. In the light-cone coordinate system the bulk metric of the ground-state of these spaces has the form:
	\begin{displaymath}
		ds^2 = r^2 \left( 2dx^{+}dx^{-} - \beta^2 r^{2} \left(dx^{+}\right)^2 + d\vec{x}^2 \right) + \frac{dr^2}{r^2}
	\end{displaymath}
where $\beta$ is some number. For the case above $\beta$ can be scaled out by a boost but it has non-trivial effects outside of the ground state. Based on the suggestions of \cite{Son:2008ye, Balasubramanian:2008dm} it later proved possible to embed these spacetimes in string theory \cite{Herzog:2008wg, Maldacena:2008wh, Adams:2008wt} although there as yet remains issues of interpretation of boundary quantities \cite{Ross:2009ar}. For other works on \Schr spactimes and non-relativistic CFTs see \cite{Goldberger:2008vg, Barbon:2008bg, Wen:2008hi, Nakayama:2008qm, Chen:2008ad, minic-2008-78, Imeroni:2008cr, PhysRevD:78:087701, Kachru:2008yh, Pal:2008rf, SekharPal:2008uy, Pal:2008id, Kovtun:2008qy, Duval:2008jg, Yamada:2008if, Lin:2008pi, Hartnoll:2008rs, Schvellinger:2008bf, Mazzucato:2008tr, Leiva:2003kd, 2000AnPhy.282..218H, Hassaine:2000ti, 2009EPJC..tmp..427Z, Ghodsi:2009hg, Gauntlett:2009zw, Donos:2009en, Donos:2009xc, Donos:2009zf}.}

{\ It seems reasonable to ask whether we can extend the fluid-gravity correspondence to discuss fluids with \Schr symmetry and perhaps shed some light on outstanding problems in non-relativistic fluid mechanics. The case of an uncharged fluid has already been considered in \cite{Rangamani:2008gi}. However, adding charge to a fluid has previously led to new conceptual insights such as the necessity of a parity violation term for a relativistic fluid, with a gravitational dual description, in a charged background \cite{Son:2009tf}. Our focus is then naturally drawn towards understanding charged fluids with \Schr symmetry. In recent papers \cite{Adams:2009dm, Imeroni:2009cs} an $U(1)$-charged, asymptotically \Schr spacetime at finite temperature has been discussed and some linear transport coefficients calculated. To obtain these results a truncation of the 10-dimensional effective Type IIB string action was isolated. Specifically we begin with a stack of D3-branes rotating in $S^{5}$ \cite{Nayak:2004rc}. The Kaluza-Klein reduction of this solution is associated with a $\RNAdS[5]$ black hole. By applying a Null Melvin Twist \cite{Alishahiha:2003ru, Gimon:2003xk}, or alternately a TsT transformation \cite{Lunin:2005jy}, to the 10-dimensional metric we induce \Schr symmetry on the boundary spacetime. Here we would like to extend the analysis of these charged, thermal systems to hydrodynamics at first order calculating all the relevant transport coefficients.}

{\ A fluid with \Schr symmetry whose gravitational dual was $(d+3)$-dimensional would occupy $d$ spatial dimensions. For comparison, an alternate approach to achieving non-relativistic symmetry is given by the ``Galilean conformal algebra'' which produces fluids moving in $d+1$ spatial dimensions. Intuitively this corresponds to permitting super-luminal communication and is achieved mathematically by taking a suitable scaling of hydrodynamic variables and spacetime coordinates. This approach preserves the form of the equation of state from the relativistic theory to the non-relativistic theory but projects out sound waves whose dispersion relations take the form $\omega \propto k$ giving an incompressible fluid \cite{Bhattacharyya:2008kq}. Unlike the \Schr algebra it does not admit a central extension; this can be interpreted as meaning the model describes gapless non-relativistic fluids. We do not consider this approach here because, as was discussed in \cite{Bhattacharyya:2008kq}, the charge conservation equation to first order reduces to demanding incompressibility. Hence the charged fluid is a trivial generalisation of the uncharged fluid under this type of scaling.}

{\ Our aim is to detail the first order corrections to a charged fluid with \Schr symmetry. We demonstrate how the general form of these corrections can be obtained by reducing conformal, relativistic currents to their non-relativistic counterparts. We then consider a five-dimensional, asymptotically \Schrcomma charged black hole spacetime and construct its fluid dual. The corrections up to first order in boundary derivatives of velocity to the metric, dilaton, gauge field and massive vector field are calculated. Finally, using the holographic dictionary in \Schr spacetimes, we shall calculate the boundary values of the stress tensor and gauge field, obtained from our charged black brane, to determine the dependence on charge and temperature profiles of the non-relativistic transport coefficients.}

{\ The paper is organised as follows. In the second section we shall discuss the theoretical formulation of non-relativistic hydrodynamics. This discussion will not be pedagogical but instead highlight a few important specifics necessary for the development of the rest of the paper, in particular, the structure of the hydrodynamic stress tensor and charge current at first order in derivatives. In the third section a map is constructed that links the transport coefficients of the relativistic charge current to their non-relativistic counterparts. This is an extension of previous work, notably, we seek to preserve the mappings of \cite{Rangamani:2008gi} between relativistic stress-energy-momentum (SEM) tensor transport coefficients and their non-relativistic counterparts. In the fourth section we shall construct an action principle for the charged, asymptotically \Schr black hole and determine its field content. For this calculation we closely follow the work of \cite{Adams:2009dm}. We then, in section five, briefly describe the method of determining the transport coefficients by solving the Einstein equations order by order in derivatives of velocity as developed in the seminal paper \cite{Bhattacharyya:2008jc}. Again the discussion of this section will avoid many of the heavier details and instead highlight the most relevant features. We refer the interested reader to \cite{Bhattacharyya:2008jc} for the minutiae. Having obtained the asymptotically \AdS metric and gauge field to first order we take the TsT of this solution to find the corresponding metric and fields with \Schr symmetry. Finally we calculate the boundary values of these fields and determine the non-relativistic transport coefficients.}

\section{Hydrodynamics}

{\ In this section we shall formulate the basics of non-relativistic hydrodynamics. Our starting point is to consider the fluid as a thermodynamic system. We then indicate how transport coefficients are related to the derivative expansions of the spatial stress tensor, energy current and charge current. In particular we discuss conventions, such as which operators to keep, in the effective field theory and set out future notation.}

{\ We shall assume our fluid is in \textbf{local} thermodynamic equilibrium in the neighbourhood of any point. The defining thermodynamic potential is the Gibbs potential:
	\begin{equation}
		G=G\left(P(x), T(x), N(x), Q^{I}(x) \right)
		\label{Eq: Hydro GibbsFreeEnergy}
	\end{equation}
where $P(x)$, $T(x)$, $N(x)$ and $Q^{I}(x)$ are the local pressure, temperature, particle number and charge of the fluid. For an isobaric and isothermal process in an open system it represents work done as particles move from one region of local equilibrium to another. Changes between equilibrium states of the system are specified by:
	\begin{equation}
		dG = \underbrace{\left(\frac{\partial G}{\partial P} \right)_{T,N,Q_{I}}}_{V(x)} dP
		   + \underbrace{\left(\frac{\partial G}{\partial T} \right)_{P,N,Q_{I}}}_{-S(x)} dT
		   + \underbrace{\left(\frac{\partial G}{\partial N} \right)_{P,T,Q_{I}}}_{\mu(x)} dN
		   + \underbrace{\left(\frac{\partial G}{\partial Q^{I}} \right)_{T,P,N}}_{\mu_{q, I}(x)} dQ^{I}
		\label{Eq: Hydro GibbsFreeEnergyVariation}
	\end{equation}
where $V(x)$, $S(x)$, $\mu(x)$ and $\mu_{q, I}(x)$ are the volume, entropy and chemical potentials respectively. It will be more profitable to consider volume densities of thermodynamic quantities which sets $G(x)$, $S(x)$, $N(x)$ and $Q^{I}(x)$ to $g(x)$, $s(x)$, $n(x)$ and $q^{I}(x)$. We shall also consider only a single $U(1)$ charge, the generalisations being clear, and hence can drop the charge group index $I$. Upon integration (\ref{Eq: Hydro GibbsFreeEnergyVariation}) yields an equation of state satisfied by the system whenever it is in equilibrium. Importantly, in a transition between equilibria, the components of $dG$ complete the specification of the thermodynamic state of the fluid. However, from a macroscopic perspective, the intensive quantities are a priori unknown.}

{\ The thermodynamic coefficients do not indicate how quantities flow between different patches of local equilibrium. This is the realm of non-relativistic hydrodynamics\footnote{Several quantities such as relativistic and non-relativistic pressure are traditionally labeled with the same symbol $P$. We shall have cause to investigate both relativistic quantities and their non-relativistic counterparts. To avoid confusion, unless an object is explicitly defined as non-relativistic, for example the mass density $\rho$, we shall adopt a notation where the subscript $_{nr}$ indicates non-relativistic while an unlabeled object is relativistic.} and as such we need to supplement our knowledge of the extensive variables with the local fluid velocity $v^{i}(x)$ and mass density $\rho(x)$. The hydrodynamic regime is characterised by four conservation equations:
	\begin{eqnarray}
		   									  \partial_{+} \rho + \partial_{i} \left( \rho v^{i} \right) &=& 0 \label{Eq: Hydro Continuity} \\
										   \partial_{+} \left(\rho v^{i} \right) + \partial_{j} \Pi^{ij} &=& 0 \label{Eq: Hydro Stresscons} \\
 		\partial_{+} \left( \epsilon_{nr} + \frac{1}{2} \rho v^2 \right) + \partial_{i} j^{i}_{\epsilon} &=& 0 \label{Eq: Hydro Energycons} \\
 														   \partial_{+} q_{nr} + \partial_{i} j^{i}_{nr} &=& 0 \label{Eq: Hydro Chargecons}
	\end{eqnarray}
where we have used $+$ to denote the time coordinate to match our later interpretation of $x^{+}$ under light-cone reduction. These equations are the continuity, momentum conservation, energy conservation and charge conservation equations respectively of the fluid. To zeroth order in derivatives of velocity and temperature we can expand the undetermined tensor objects, $\Pi_{ij}$, $j^{i}_{\epsilon}$ and $j^{i}_{nr}$ as:
	\begin{eqnarray}
					     \Pi^{ij} &=& \rho v^{i} v^{j} + P_{nr} \delta^{ij} \label{Eq: Hydro Stresstensor0} \\
				 j^{i}_{\epsilon} &=& \left(\epsilon_{nr} + P_{nr} + \frac{1}{2} \rho v^2\right) v^{i} \label{Eq: Hydro Energycurrent0} \\
				 	j^{i}_{nr}    &=& q_{nr} v^{i} \label{Eq: Hydro Charge3current0}
	\end{eqnarray}
where $\epsilon_{nr}$ and $P_{nr}$ are the fluid's energy density and pressure. We shall call (\ref{Eq: Hydro Stresstensor0}), (\ref{Eq: Hydro Energycurrent0}) and (\ref{Eq: Hydro Charge3current0}) the stress tensor, energy density current and charge density current at zeroth order. These quantities characterise a perfect fluid which does not lose energy due to internal friction as it has no viscosity.}

{\ Following \cite{Rangamani:2009xk} we consider adding terms to our spatial stress tensor and currents with single derivatives of velocity. Our undetermined tensor quantities in (\ref{Eq: Hydro Stresscons}) and (\ref{Eq: Hydro Energycons}) take the form:
	\begin{eqnarray}
						\Pi^{ij} &=& \rho v^{i} v^{j} + P_{nr} \delta^{ij} - \eta_{nr} \sigma^{ij} - \zeta_{nr} \theta \delta^{ij} \label{Eq: Hydro StresstensorI} \\
				j^{i}_{\epsilon} &=& \left(\epsilon_{nr} + P_{nr} + \frac{1}{2} \rho v^2\right) v^{i} - \eta_{nr} \sigma^{ij} v_{j} - \kappa_{T} \delta^{ij} \partial_{j} T - \nonumber \\
								&\;& - \varpi \delta^{ij} \partial_{j} \ln\left[ 
									 \frac{r_{+}\left(T,\frac{\mu}{T},\frac{\mu_{q}}{T}\right) \left[r_{+}^{4}\left(T,\frac{\mu}{T},\frac{\mu_{q}}{T}\right) -\frac{8}{3} \frac{\mu_{q}^2}{\mu} r_{+}^{2} \left(T,\frac{\mu}{T},\frac{\mu_{q}}{T}\right) \right]^{\frac{1}{4}} }
									 {\left(r_{+}^2\left(T,\frac{\mu}{T},\frac{\mu_{q}}{T}\right)+\frac{4}{3} \frac{\mu_{q}^2}{\mu} \right)} \right] 
									 \label{Eq: Hydro EnergycurrentI}
	\end{eqnarray}
where we have assumed a flat background. The stress tensor $\Pi_{ij}$ is unchanged from the uncharged case of \cite{Rangamani:2008gi} but the energy current $j^{i}_{\epsilon}$ has received an additional correction which vanishes when the charge is set to zero. In these expressions we have used the following quantities:
	\begin{eqnarray}
		  \theta      &=& \partial_{i} v^{i} \nonumber \\
		  \sigma^{ij} &=& \left(\partial^{i} v^{j} + \partial^{j} v^{i} - \frac{2 \delta^{i j}}{d} \theta \right) \nonumber \\
		 r_{+}(x,y,z) &=& \frac{\pi}{2\sqrt{2}} \left(-\frac{x}{y}\right)^{\frac{1}{2}} \left[1 + \sqrt{1 + \frac{32}{3} \frac{z^2}{\pi^2}} \right] \label{Eq: Hydro Defofr}
	\end{eqnarray}
assuming our non-relativistic fluid occupies $d$ spatial dimensions. The coefficients of (\ref{Eq: Hydro StresstensorI}) and all but the last coefficient of (\ref{Eq: Hydro EnergycurrentI}) have standard physical interpretations; $\eta_{nr}$ represents the fluid's shear viscosity, $\zeta_{nr}$ the bulk viscosity and $\kappa_{T}$ the thermal conductivity \cite{LandauLifschitz:1284488}. We shall call the new coefficient, $\varpi$, the contribution to the energy current from charge.}

{\ As regards determining the first order corrections to the current vector we note that at zeroth order the conservation equations (\ref{Eq: Hydro Continuity})-(\ref{Eq: Hydro Chargecons}) can be written as:
	\begin{eqnarray}
								  \partial_{+} \rho + v^{i} \partial_{i} \rho + \rho \partial_{i} v^{i} &=& 0 \nonumber \\
					  \partial_{+} v^{i} + v^{j}\partial_{j} v^{i} + \frac{1}{\rho} \partial^{i} P_{nr} &=& 0 \nonumber \\
		\partial_{+} \epsilon_{nr} + \partial_{i} \left(\epsilon_{nr} v^{i}\right) + P_{nr} \partial_{j} v^{j} &=& 0 \nonumber \\
							\partial_{+} q_{nr} + v^{i} \partial_{i} q_{nr} + q_{nr} \partial_{i} v^{i} &=& 0 \nonumber
	\end{eqnarray}
where the equation of state for the fluid relates $\epsilon_{nr}$ and $P_{nr}$. Hence, if we obtain a complete solution to the above equations they allow us to replace time derivatives of the variables $\epsilon_{nr}$, $\rho$, $q_{nr}$ and $v^{i}$ for spatial derivatives at first order making an error in our final results that, overall, is second order in derivatives and can therefore be ignored. We can thus write our non-relativistic current vector as:
	\begin{equation}
		j^{i}_{nr} =  q_{nr} v^{i} - \kappa_{nr} \delta^{ij} \partial_{j} q_{nr} - \gamma^{ij}_{nr} \partial_{j} \epsilon_{nr} - \digamma^{ij}_{nr} \partial_{j} \rho
					  - \mho_{nr} \left[ \epsilon^{ij} v^{k} \left( \partial_{j} v_{k} - \partial_{k} v_{j} \right) + v^{i} \epsilon^{jk} \partial_{j} v_{k} \right] \label{Eq: Hydro ChargecurrentI}
	\end{equation}
Here $\kappa_{nr}$ is the non-relativistic diffusion constant and $\mho_{nr}$ the parity violation coefficient. We shall call the tensor objects $\gamma^{ij}_{nr}$ and $\digamma^{ij}_{nr}$ the contributions of energy density and mass density to the charge current respectively.}

\section{Light-cone reduction of charged relativistic fluids}

{\ One way of obtaining the form of the first order corrections to (\ref{Eq: Hydro Stresstensor0}), (\ref{Eq: Hydro Energycurrent0}) and (\ref{Eq: Hydro Charge3current0}) is to light-cone reduce the SEM tensor and charge current of a relativistic fluid. In particular we shall consider a conformal, relativistic fluid and then the light-cone reduction will lead to a hydrodynamic system with \Schr symmetry \cite{Rangamani:2008gi}. We begin this section by stating the conformally invariant, relativistic conserved currents, specifically, our choice of variables. We then summarise the previous work of \cite{Rangamani:2008gi} on the uncharged fluid before determining a map between relativistic charge coefficients and their non-relativistic counterparts.}

{\ Determining the relativistic conserved currents provides the starting point for our analysis. Using conformality, the choice of Landau frame and relativistic invariance, as shown in \cite{Rangamani:2009xk}, it is possible to decompose these currents to first order in derivatives in the following manner:
	\begin{eqnarray}
		T^{\mu \nu} &=& \left(\epsilon+P\right) u^{\mu} u^{\nu} + P \eta^{\mu \nu} 
						- 2 \eta \left( P^{\mu \alpha} P^{\nu \beta} \nabla_{\left( \alpha \right.} u_{\left. \beta \right)} - \frac{1}{\dimS} \nabla_{\alpha} u^{\alpha} P^{\mu \nu} \right) \label{Eq: LC StressTensor} \\
			j^{\mu} &=& q u^{\mu} - \kappa_{q} P^{\mu \nu} \nabla_{\nu} q - \gamma P^{\mu \nu} \nabla_{\nu} \epsilon 
							- \mho \epsilon_{\alpha \beta \gamma}\;^{\mu} u^{\alpha} \nabla^{\beta} u^{\gamma} \label{Eq: LC Chargecurrent}
	\end{eqnarray}
where $P^{\mu \nu}=u^{\mu} u^{\nu} + \eta^{\mu \nu}$. The decomposition in (\ref{Eq: LC Chargecurrent}) is not explicitly conformal in $\kappa_{q}$ and $\gamma$ and, therefore, they must satisfy the following relation:
	\begin{equation}
		\dimS \kappa_{q} q + \dimB \gamma \epsilon =0
		\label{Eq: LC Scalingrelation}
	\end{equation}
The final term of (\ref{Eq: LC Chargecurrent}) is specific to four dimensions as this is the only case where there is a first order parity violating contribution.}

{\ The parity violating term of (\ref{Eq: LC Chargecurrent}) may at first be alarming. Traditional approaches to determining the corrections to $J^{\mu}$ such as Israel-Stewart theory \cite{Israel:1976tn, Israel:1979wp} typically ignore or argue this coefficient away. However, as was shown in \cite{Son:2009tf}, in a theory with a charged background, consideration of triangle anomalies leads to its presence. By demanding positivity of the entropy current it is possible to find an expression for $\mho$ in terms of the gauge anomaly coefficients and the hydrodynamic variables. We refer the reader to \cite{Son:2009tf} for the general case and shall be content here to determine $\mho$ for the specific case of a fluid dual to an asymptotically \Schrcomma $U(1)$-charged black hole. We shall find it to be fixed by the charge $q$ and the fluid pressure $P$ up to an anomaly dependent constant. The other terms in (\ref{Eq: LC Chargecurrent}) have more sedate interpretations as the charge diffusion coefficients $\kappa_{q}$ and contributions of the energy density to the charge current $\gamma$.}

{\ As was demonstrated in \cite{Rangamani:2008gi} at first-order in derivatives of fluid velocity there exists a map between the relativistic SEM tensor variables, $(u^{\mu}, \epsilon,P,\eta)$, and non-relativistic $(v^{i}, \rho, \epsilon_{nr},P_{nr},\eta_{nr})$ variables. We shall seek to maintain these relations and augment them with our charge variable maps. In particular we can use them in our reduction of the charge current. Summarising the results of \cite{Rangamani:2008gi} we begin by assuming that our fluid lives on a Minkowskian background with metric:
	\begin{equation}
		ds^{2} = 2dx^{+} dx^{-} + d\vec{x}^2
		\label{Eq: LC BoundaryMetric}
	\end{equation}
and that the relativistic hydrodynamic variables and velocities depend only trivially on the $x^{-}$ direction. This suggests we make the following identifications:
	\begin{eqnarray}
		T^{++} &=& \rho \nonumber \\
		T^{+i} &=& \rho v^{i} \nonumber \\
		T^{+-} &=& -\left( \epsilon_{nr} + \frac{1}{2} \rho v^{2} \right) \label{Eq: LC TensorComplexIdentification} \\
		T^{-i} &=& -j^{i}_{\epsilon} \nonumber \\
		T^{ij} &=& \Pi^{ij} \nonumber
	\end{eqnarray}
which come from comparing SEM tensor conservation equations in our choice of coordinates and (\ref{Eq: Hydro Continuity})-(\ref{Eq: Hydro Energycons}). The $T^{++}$ component of the SEM tensor implies the following identification between variables:
	\begin{eqnarray}
		\rho = \left(\epsilon + P \right) \left(u^{+}\right)^2
		\label{Eq: LCR Unchargedmap}
	\end{eqnarray}
while the $T^{+i}$ component indicates that:
	\begin{displaymath}
		v^{i} = \frac{u^{i}}{u^{+}} - \frac{\eta}{\rho} \left( \partial_{i} u^{+} - \frac{u^{+}}{\left(\epsilon+P\right)} \partial_{i} P  \right)
	\end{displaymath}
The other quantities of importance we list for completeness:
	\begin{eqnarray}
			      		 \Pi^{ij} &=& \rho v^{i} v^{j} + P \delta^{ij} - \eta u^{+} \sigma^{ij} \label{Eq: LC NRStressTensor} \\
						   P_{nr} &=& P \label{Eq: LC NRPressure} \\
		       			\eta_{nr} &=& \eta u^{+} \\
			 		\epsilon_{nr} &=& \frac{1}{2} \left(\epsilon-P \right) \\
				 j^{i}_{\epsilon} &=& \left( \epsilon_{nr} + P_{nr} + \frac{1}{2} \rho v^2 \right) v^{i} - \eta_{nr} \sigma^{ij} v_{j} 
									  - \frac{2 \eta_{nr} P_{nr}}{\rho} \delta^{ij} \left[ \frac{3}{2} \frac{\partial_{j} \epsilon_{nr}}{\epsilon_{nr}} - \frac{\partial_{j} \rho}{\rho} \right] \label{Eq: LC Energycurrent}
	\end{eqnarray}
where (\ref{Eq: LC NRStressTensor}) has no bulk viscosity term as conformal invariance of the parent relativistic theory set this to zero. For a conformal relativistic fluid in $\dimB$ spacetime dimensions the equation of state is supplied by tracelessness of the SEM tensor and implies $\epsilon=\dimS P$ and therefore, using the above maps, the non-relativistic fluid satisfies $\epsilon_{nr}=\frac{d}{2}P_{nr}$.}

{\ In \cite{Rangamani:2008gi} the final two terms of (\ref{Eq: LC Energycurrent}) are eliminated in favour of $\partial_{i}T$ and the resultant coefficient is interpreted as the thermal conductivity $\kappa_{T}$. This can be done because the equation of state for an uncharged fluid is given by:
	\begin{displaymath}
		\epsilon_{nr} = \alpha \left( \frac{T^2}{\mu} \right)^{\frac{d+2}{2}}
	\end{displaymath}
where $\alpha$ is a constant as detailed in \cite{Kovtun:2008qy}. However for a charged fluid there is an additional scale in the problem, $\mu_{q}$, and therefore our equation of state takes the more generic form: $P_{nr}=T^{\frac{d+2}{2}}g\left(\frac{\mu}{T}, \frac{\mu_{q}}{T}\right)$ which prevents us from eliminating $\partial_{i} \epsilon_{nr}$ and $\partial_{i} \rho$ completely. We shall return to the interpretation of these terms in section 5.}

{\ Considering now the charge density current it is clear that it satisfies the conservation equation:
	\begin{displaymath}
		\partial_{+} j^{+} + \partial_{i} j^{i} = 0
	\end{displaymath}
Using the expression for $j^{\mu}$ from (\ref{Eq: LC Chargecurrent}) and the maps (\ref{Eq: LCR Unchargedmap})-(\ref{Eq: LC Energycurrent}) we find that $j^{+}$ has the form:
	\begin{eqnarray}
		j^{+} = q u^{+} + \mho (u^{+})^2 \partial^{j} v^{k} \epsilon_{jk} \label{Eq: LC TemporalPart}
	\end{eqnarray}
where we have used the scaling relation (\ref{Eq: LC Scalingrelation}) to annihilate the term proportional to $\theta$ and set $\epsilon_{+-ij}=-\epsilon_{ij}$. It is a satisfying occurrence that the only correction to the identification $j^{+}=q_{nr}$ at first order is a piece which accounts for the anomalies in the relativistic theory. Indeed, in the holographic model we shall construct in future sections, if the Chern-Simon's coupling in our action is set to zero, then $q_{nr}$ is just a scaling by $u^{+}$ of $q$. Reducing the spatial part of the current leads to:
	\begin{eqnarray}
		j^{i} &=& j^{+} v^{i} - \frac{\kappa_{q}}{u^{+}} \partial^{i} j^{+} 
				  - \mho (u^{+})^2 \left[ \epsilon^{ij} v^{k} \left( \partial_{j} v_{k} - \partial_{k} v_{j} \right) + v^{i} \epsilon^{jk} \partial_{j} v_{k}  \right] - \nonumber \\
			 		&\;& - \left(\frac{d}{d+2} \right) \left[ j^{+} \left(\frac{\left(d+4\right)}{d} \frac{\eta u^{+}}{\rho} - \frac{\kappa_{q} }{u^{+}} \right) \delta^{ij} + \mho \epsilon^{ij} \right] \left( \frac{\partial_{j} \epsilon}{2\epsilon} \right) + \nonumber \\
			    &\;& + \left[ \left(\frac{\kappa_{q}}{u^{+}}+\frac{\eta u^{+}}{\rho}\right) \delta^{ij} + \mho \epsilon^{ij} \right] \left( \frac{\partial_{j} \rho}{2 \rho} \right) \nonumber 
	\end{eqnarray}
Additionally, in light of the above expansion, it seems reasonable to define $\kappa_{nr}=\frac{\kappa_{q}}{u^{+}}$ and $\mho_{nr}=\mho (u^{+})^2$. The remaining terms, proportional to $\partial_{i} P$ and $\partial_{i} \rho$, merit further consideration for the same reasons as the final terms of (\ref{Eq: LC Energycurrent}) and we shall return to them in light of section 5. However, using the equation of state to find $\epsilon$ in terms of $\epsilon_{nr}$ the spatial part of the current can be rewritten as:
	\begin{eqnarray}
		j^{i} &=& q_{nr} v^{i} - \kappa_{nr} \partial^{i} q_{nr} 
				  - \mho_{nr} \left[ \epsilon^{ij} v^{k} \left( \partial_{j} v_{k} - \partial_{k} v_{j} \right) + v^{i} \epsilon^{jk} \partial_{j} v_{k}  \right] - \nonumber \\
			 &\;& - \left(\frac{d}{d+2} \right) \left[ q_{nr} \left(\frac{\left(d+4\right)}{d} \frac{\eta_{nr}}{\rho} - \kappa_{nr} \right) \delta^{ij} + \frac{2\left(d+2\right)}{d} \frac{\mho_{nr} \epsilon_{nr}}{\rho} \epsilon^{ij} \right] 
				  \left( \frac{\partial_{j} \epsilon_{nr}}{2\epsilon_{nr}} \right) + \nonumber \\
			 &\;& + \left[ q_{nr} \left(\kappa_{nr}+\frac{\eta_{nr}}{\rho}\right) \delta^{ij} + \frac{2\left(d+2\right)}{d} \frac{\mho_{nr} \epsilon_{nr}}{\rho} \epsilon^{ij} \right] \left( \frac{\partial_{j} \rho}{2 \rho} \right) \nonumber \label{Eq: LC SpatialPart}
	\end{eqnarray}
which represents the maximal possible reduction into our chosen non-relativistic operators. This matches (\ref{Eq: Hydro ChargecurrentI}) if we identify the coefficient of $\partial_{j} \epsilon_{nr}$ with $\gamma^{ij}_{nr}$ and that of $\partial_{j} \rho$ with $-\digamma^{ij}_{nr}$.}

{\ So far we have only related the non-relativistic and relativistic variables. A priori we have no knowledge of the functional form of the relativistic charge coefficients in terms of local charge and temperature profiles. Hence we also do not know the functional forms of the non-relativistic coefficients in terms of the corresponding non-relativistic charge and temperature. In the next section we specialise our discussion to one particular fluid; a charged, \Schr invariant fluid which has as its dual a charged black hole with asymptotically \Schr boundary. The \AdSnrCFT will then readily allow us to calculate these dependencies; an otherwise difficult problem for a general fluid.}

\section{Charged, asymptotically Schr\"{o}dinger black hole construction}

{\ As our focus is now on \Schr fluids with gravity duals we can apply the \AdSnrCFT \cite{Herzog:2008wg, Maldacena:2008wh, Balasubramanian:1999re} to our investigation. This provides a map from an asymptotically \Schr space-time to a boundary field theory with \Schr symmetry and hence allows us to determine expectation values of the spatial stress tensor, energy current and charge current at zeroth order. In particular, we begin with an uplift for $\RNAdS[5]$ which we can TsT transform to give the bulk fields boundary \Schr symmetry. To calculate the temperature and charge profiles corresponding to our dual black hole we will need to construct a suitable action yielding these bulk fields. We lean heavily on the formalism of \cite{Adams:2009dm} and find a 10-dimensional action from an effective description of Type IIB string theory whose Kaluza-Klein reduction leads to a five dimensional theory with the correct field content. It turns out that this reduced action contains only four fundamental fields; the gravitational field, the dilaton, an $R$-charged gauge field and a massive vector that is now standard fare in \Schr spacetimes \cite{Son:2008ye, Herzog:2008wg, Maldacena:2008wh, Adams:2008wt}. We complete this section by calculating the thermodynamics of our bulk spacetime, in particular, the specific heat at constant pressure, particle number and charge.}

{\ We shall take $m$ to be the mass of our black hole and $Q$ to be its charge. From \cite{Buchel:2006gb} a suitable $\RNAdS[5]$ ansatz is:
	\begin{eqnarray}
		ds^2_{10} &=& -r^2 f(m,Q,r) dt^2 +\frac{dr^2}{r^2 f(m,Q,r)} + r^2\left(dx^{2} + dy^2 + dz^2 \right) + \nonumber \\
			     &\;& + \left(d\psi + \mathcal{A}_{(1)} - \frac{2}{\sqrt{3}}  A_{Q} \right)^2 + d\Sigma_{4}^{2} \label{Eq: CSBH Ansatzmetric} \\
		  F_{(5)} &=& 2 (1+*_{10})\left[\left(d\psi + \mathcal{A}_{(1)} - \frac{2}{\sqrt{3}}  A_{Q} \right) \wedge J_{(2)} - \frac{1}{\sqrt{3}} *_{5} F_{Q} \right] \wedge J_{(2)} \label{Eq: CSBH 5Form} \\
		  	A_{Q} &=& \frac{\sqrt{3}}{2} \frac{Q}{r^2} dt \label{Eq: CSBH Ansatzgauge} \\
	    	F_{Q} &=& d A_{Q} \\
		  f(m,Q,r)&=& 1- \frac{m}{r^{4}}+\frac{Q^2}{r^{6}}  \label{Eq: CSBH Defnoff}
	\end{eqnarray}
where $*_{5}$ and $*_{10}$ are the five- and ten-dimensional Hodge stars defined in the appendix and the unusual factors of $-\frac{2}{\sqrt{3}}$ are present so that our $A_{Q}$ matches that of \cite{Banerjee:2008th}. Note that by choosing $A_{Q} = a(r) dt$ we only have an electric field turned on. The final two terms of the ten-dimensional metric are the five-dimensional metric on the unit $S^{5}$ given by a $U(1)$ fibration over $\mathbb{CP}^2$ with a gravi-photon $a(r) dt$ turned on. The one-form $\mathcal{A}_{(1)}$ and the two form $J_{(2)}$ are called the K\"{a}hler potential and form respectively and are defined by the following equations:
	\begin{eqnarray}
							  J_{(2)} &=& \frac{1}{2} d\mathcal{A}_{(1)} \nonumber \\
		\mathrm{Vol}(\mathbb{CP}^{2}) &=& \frac{1}{2} J_{(2)} \wedge J_{(2)} \nonumber
	\end{eqnarray}
The fields of (\ref{Eq: CSBH Ansatzmetric})-(\ref{Eq: CSBH Defnoff}) satisfy the equations of motion from the following truncation of Type IIB string theory:
	\begin{equation}
		S_{10} = \frac{1}{16 \pi G_{10}} \int \left[ (*_{10}1) e^{-2\Phi} \left(R^{(10)} + 4 (\partial \Phi)^2 \right) -\frac{1}{4} F_{(5)} \wedge *_{10} F_{(5)} \right] \label{Eq: CSBH 10daction}
	\end{equation}
where we have chosen the string frame. As usual, self-duality of the RR 5-form must be imposed after variation to complete the specification of the equations of motion.}

{\ Using the five-form Bianchi identify $dF_{(5)}=0$ we find our choice for the decomposition of the RR 5-form implies that $F_{Q}$ satisfies the following equation of motion:
	\begin{displaymath}
		d *_{5} F_{Q} - 4 \kappa_{CS} F_{Q} \wedge F_{Q} = 0
	\end{displaymath}
where $\kappa_{CS}=-\frac{1}{2\sqrt{3}}$ is the Chern-Simon's coupling of the gauge field $A_{Q}$. This follows from an action of the form:
	\begin{displaymath}
		\int \left[ F_{Q} \wedge *_{5} F_{Q} - \frac{8}{3} \kappa_{CS} A_{Q} \wedge F_{Q} \wedge F_{Q} \right]
	\end{displaymath}
which, upon compactification, we shall use to substitute for the $F_{(5)}$ terms of (\ref{Eq: CSBH 10daction}). There is one caveat in this replacement and it is that the cosmological constant in the five-dimensional theory receives contributions from the volume form pieces of (\ref{Eq: CSBH 5Form}). The correct cosmological constant can be determined by examining the equations of motion.}

{\ Also, we now generalise to arbitrary $\kappa_{CS}$ which will clarify the role of the Chern-Simon's term in generating the parity violating coefficient $\mho$ of (\ref{Eq: LC Chargecurrent}). Compactifying $S^{5}$ we can set the resultant scalar field associated with the metric, $\sigma$, to be equal to the dilaton $\Phi$. Further, we find $\Phi=0$ is a consistent solution to the equations of motion and hence our action (\ref{Eq: CSBH 10daction}) reduces to:
	\begin{equation}
		S_{5} = \frac{1}{16 \pi G_{5}} \int \left[ \vol{M} \left(R^{(5)} + 12 \right) - 2 F_{Q} \wedge *_{5} F_{Q} 
							+ \frac{16\kappa_{CS}}{3} A_{Q} \wedge F_{Q} \wedge F_{Q} \right] \label{Eq: CSBH TypeIIBRelReduction}
	\end{equation}
in the Einstein frame where $G_{5}=\frac{G_{10}}{\mathrm{vol}\left(S^{5}\right)}$. This matches the Einstein-Maxwell action used to calculate the fluid metric up to second order in \cite{Banerjee:2008th}.}

{\ We would now like to induce \Schr symmetry on the boundary of the $\RNAdS[5]$ spacetime given by (\ref{Eq: CSBH Ansatzmetric})-(\ref{Eq: CSBH Defnoff}). To do this we can apply one of a pair of solution generating techniques at the level of the equations of motion; the Null-Melvin twist \cite{Alishahiha:2003ru, Gimon:2003xk} or TsT \cite{Lunin:2005jy}. They both take an $\RNAdS[5] \times \chi^{5}$ manifold where $\chi^{5}$ is Sasaki-Einstein and yield an asymptotically \Schrcomma charged black hole with a deformed $\chi^{5}$. The former, NMT, begins by boosting our solution along one of the spatial isometry directions, say $y$ by a rapidity of $\gamma$. Two T-dualisations along $y$ are then performed with a twist of the one-form $d\psi \rightarrow \upsilon d\psi$ sandwiched between them. We then boost the resultant fields by $-\gamma$. Finally $\upsilon$ is scaled to zero and $\gamma \rightarrow \infty$ while keeping $\beta = \frac{1}{2} \upsilon e^{\gamma}$ constant. The latter technique, TsT, involves a twist in the $x^{-}=\frac{1}{2\beta}(y-t)$ direction by $\alpha d\psi$ between two T-dualisations along the $\psi$ direction. This second technique can be applied to any spacetime with a $U(1) \times U(1)$ isometry. Moreover, when one of the $U(1)$ isometries is null it can be shown that the two techniques coincide \cite{Rangamani:2008gi}.}

{\ Applying a TsT along the Hopf direction $\psi$ as discussed in the appendix, where we also give the heavier details of our notation, leads to the following fields:
	\begin{eqnarray}
		  (ds_{10}^2)'' &=& \frac{r^2}{k} \left[ -\beta^2 r^2 f(m,Q,r) (dt+dy)^2 - f(m,Q,r)dt^2 +dy^2 + kd\vec{x}^2 \right] + \nonumber \\
			 		   			 &\;& + \frac{dr^2}{r^2 f(m,Q,r)} + \frac{\left(d\psi + \mathcal{A}_{(1)} - \frac{2}{\sqrt{3}} A_{Q}\right)^2}{k} + d\Sigma_{4}^{2} \label{Eq: CSBH Backgroundmetric} \\
	    B''_{(2)} 		&=& \frac{\beta r^2}{k} \left(dy + f(m,Q,r) dt \right) \wedge \left(d\psi + \mathcal{A}_{(1)} - \frac{2}{\sqrt{3}} A_{Q} \right) \\
	    F''_{(3)} 		&=& -\frac{2Q\beta}{r^3} J_{(2)} \wedge dr \\
	    F''_{(5)} 		&=& F_{(5)} + B''_{(2)} \wedge F''_{(3)} \\
		 \exp(2\Phi'')  &=& \frac{1}{k} \label{Eq: CSBH Backgrounddilaton}
	\end{eqnarray}
where we have set $\alpha = 1$ so that boost and twist parameters of NMT and TsT respectively coincide. We have defined:
	\begin{equation}
		k =	1 + \beta^2 r^2 (1-f(m,Q,r))
		\label{Eq: CSBH Defofk}
	\end{equation}
in correspondence with the notation of \cite{Herzog:2008wg} and note that our results are in agreement with \cite{Adams:2009dm}. The operation $*''_{10}$ is the Hodge dual in the Melvinised spacetime and its definition is also detailed in the appendix.}

{\ The self-duality of the 5-form $F''_{(5)}$ allows us to determine a relation between $*_{5} F_{Q}$ in the un-Melvinised spacetime and the quantities $\Phi''$, $f$, $A_{M}$ and $F_{Q}$ in the Melvinised spacetime\footnote{See appendices for further details.}. Hence we can write:
	\begin{eqnarray}
		S_{5} &=& \frac{1}{16 \pi G_{5}} \int \left[ \vol{M''} e^{-2\Phi} \left(R^{(5)} + 16 - 4 e^{2\Phi} \right) - 4 e^{\Phi} F \wedge *''_{5} F - \frac{1}{2} e^{-3\Phi} F_{M} \wedge *''_{5} F_{M}  - \right. \nonumber \\ 
			 &\;&	\phantom{\frac{1}{16 \pi G_{5}} \int \left[ \right. } - 4e^{-\Phi} A_{M} \wedge *''_{5} A_{M} - \frac{2}{3} e^{-\Phi} \left( A_{M} \wedge F_{Q} \right) \wedge *''_{5} \left( A_{M} \wedge F_{Q} \right) - \nonumber \\
			 &\;& \phantom{\frac{1}{16 \pi G_{5}} \int \left[ \right. }  - 4 e^{-\Phi} \left(\frac{1}{\sqrt{3}} F_{Q} + F \wedge A_{M} \right) \wedge *''_{5} \left(\frac{1}{\sqrt{3}} F_{Q} + F \wedge A_{M} \right) - \nonumber \\ 
			 &\;& \left. \phantom{\frac{1}{16 \pi G_{5}} \int \left[ \right. } - \frac{2}{3} e^{\Phi} F_{Q} \wedge *''_{5} F_{Q} + \frac{16 \kappa_{CS}}{3} A_{Q} \wedge F_{Q} \wedge F_{Q} \right] \label{Eq: CSBH SchrAction}
	\end{eqnarray}
where, as was discovered in \cite{Adams:2009dm}, the equation of motion for the $F$ field is completely algebraic:
	\begin{displaymath}
		F = e^{-2\Phi} *''_{5} \left( A_{M} \wedge *''_{5} \left(\frac{1}{\sqrt{3}} F_{Q} + F \wedge A_{M} \right) \right)
	\end{displaymath}
and it is therefore only an auxiliary field which is merely present to simplify our action.}

{\ Considering (\ref{Eq: CSBH SchrAction}) we can see that the Kaluza-Klein reduction of the fields (\ref{Eq: CSBH Backgroundmetric})-(\ref{Eq: CSBH Backgrounddilaton}) in the string frame is:
	\begin{eqnarray}
		(ds_{5}^2)'' &=& \frac{r^2}{k} \left[ -\beta^2 r^2 f(m,Q,r) (dt+dy)^2 - f(m,Q,r) dt^2 +dy^2 + kd\vec{x}^2 \right] + \nonumber \\
								&\;& + \frac{dr^2}{r^2 f(m,Q,r)} \label{Eq: CSBH 5dmetric} \\
		   	  A_{Q}  &=& \frac{\sqrt{3}}{2} \frac{Q}{r^2} dt \\
	   		  A_{M}  &=& \frac{\beta r^2}{k} \left(dy + f(m,Q,r) dt \right) \label{Eq: CSBH MassiveGauge}
	\end{eqnarray}
where the massless one form and non-trivial dilaton field come from the metric while the massive vector field originates in the NS-NS two form $B_{(2)}$. We note that the existence of a charged, massless one-form whose boundary value shall be interpreted as sourcing the charge of our fluid and a massive vector field which lacks a corresponding conserved current on the boundary.}

{\ With the full 5-dimensional zeroth order metric (\ref{Eq: CSBH 5dmetric}) available we can now determine the thermodynamics of our fluid. Given $m$ and $Q$, our metric has a horizon whenever $f(m,Q,r)=0$ and hence will have an associated Hawking temperature $T$. We take $r_{+}$ to be the location of the outermost horizon which turns out to be Killing as our spacetime is stationary. The temperature can be found by determining the surface gravity, $\kappa$, from the following formula:
	\begin{equation}
		\kappa^2 = - \left. \frac{1}{2} \left( \nabla_{\mu} \xi_{\nu} \right) \left( \nabla^{\mu} \xi^{\nu} \right) \right|_{r=r_{+}}
		\label{Eq: CSBH SurfaceGrav}
	\end{equation}
where $\xi^{a}$ is the null generator associated with the horizon. We are working in the Einstein frame attained from (\ref{Eq: CSBH 5dmetric}) by conformally rescaling the metric with the dilaton:
	\begin{equation}
		ds^{2}_{E} = e^{-\frac{2}{3} \Phi''} (ds_{5}^2)''
		\label{Eq: CSBH 5dmetricEins}
	\end{equation}
For (\ref{Eq: CSBH 5dmetricEins}) the (Killing) vector generating the horizon is proportional to $\left(\partial_{t} \right)^{a}$. To determine the constant of proportionality we note that in lightcone coordinates, $x^{+}$ and $x^{-}$, $\left(\partial_{+}\right)^{a}$ is the generator of boundary time translations and hence if we set its coefficient to be unit we find:
	\begin{eqnarray}
		\xi^{a} &=& \frac{1}{\beta} \left(\partial_{t} \right)^{a} \nonumber \\
					  &=& \left(\partial_{+}\right)^{a} - \frac{1}{2\beta^2} \left(\partial_{-}\right)^{a} \label{Eq: CSBH NullKilling}
	\end{eqnarray}
Substituting this result into (\ref{Eq: CSBH SurfaceGrav}) the temperature, $T=\frac{\kappa}{2\pi}$, is:
	\begin{equation}
		T = \frac{r_{+}}{2 \pi \beta} \left( 2 - \frac{Q^2}{r_{+}^{6}} \right)
		\label{Eq: CSBH HawkingTemp}
	\end{equation}
In the limit that $Q \rightarrow 0$ this coincides with the result of \cite{Rangamani:2008gi}. Of note, the temperature of the relativistic precursor theory (\ref{Eq: CSBH Ansatzmetric})-(\ref{Eq: CSBH Defnoff}) is given in terms of the non-relativistic temperature (\ref{Eq: CSBH HawkingTemp}) by $\beta T$.}

{\ The entropy associated with (\ref{Eq: CSBH 5dmetric}) can be calculated using the Hawking formula:
	\begin{displaymath}
		S = \frac{A}{4 G_{5}}	
	\end{displaymath}
where $A$ is the area of the event horizon at $r=r_{+}$. The solution turns out to be independent of $\beta$ which is to do with the fact that our metric was generated by a series of boosts and dualities from (\ref{Eq: CSBH Ansatzmetric}) as discussed in \cite{Gimon:2003xk}. For our solution, taking $V_{3}$ to be volume of the horizon, we find the entropy in the boundary is given by:
	\begin{equation}
		S = \frac{r_{+}^{3}}{4 G_{5}} V_{3}
		\label{Eq: CSBH HawkingEntropyDensity}
	\end{equation}
whose form in terms of $r_{+}$ is unchanged from the uncharged case discussed in \cite{Rangamani:2008gi}.}

{\ In (\ref{Eq: CSBH NullKilling}) we have a generator of time translation on the boundary $\left(\partial_{+}\right)^{a}$ and an additional generator $\left(\partial_{-}\right)^{a}$. The former corresponds to the Hamiltonian $\hat{H}$ in the dual field theory while the latter should be interpreted as the particle number generator $\hat{N}$. Hence the coefficient of $\left(\partial_{-}\right)^{a}$ in the null generator $\xi^{a}$ is the particle number chemical potential:
	\begin{displaymath}
		\mu = - \frac{1}{2\beta^2}
	\end{displaymath}
However, this is not the only chemical potential in the thermodynamics of our solution as the charge can also vary. This charge chemical potential can be found from the asymptotic values of the ``boundary time'' component of the gauge field and is given by:
	\begin{eqnarray}
		\mu_{q} &=& A_{+}(r_{+})-A_{+}(\infty) \nonumber \\
				&=& \frac{\sqrt{3}Q}{4 \beta r_{+}^{2}} \nonumber \\
				&=& \frac{\sqrt{6}}{4} \frac{Q \left(-\mu\right)^{\frac{1}{2}}} {r_{+}^{2}} \label{Eq: CSBH ChemicalPotential}
	\end{eqnarray}
Note that we have explicitly defined both the chemical potentials as non-relativistic quantities.}

{\ We now have sufficient information to specify our density matrix and can therefore obtain the thermodynamic potential of our ensemble. The existence of two chemical potentials implies that the density matrix has the following form:
	\begin{displaymath} 
		\hat{\rho} = \exp \left[ - \left(\frac{\hat{H} - \mu \hat{\partial}_{-} - \mu_{q} \hat{J}^{+}}{T} \right) \right]
	\end{displaymath}
The trace of the density matrix gives us the partition function whence the thermodynamic potential is determined by:
	\begin{equation}
		\tilde{G}(V_{2},T,\mu,\mu_{q}) = - T \ln \Xi(T,\mu,\mu_{q})
		\label{Eq: CSBH GeneralGibbs}
	\end{equation}
where $\Xi=\mathrm{tr}\left(\hat{\rho}\right)$ and $V_{2}$ is the two-dimensional spatial volume of the \Schr theory. This free energy represents the work done by the system when the chemical potential difference between two neighbouring regions of equilibrium changes in an isothermal and isochoric process. The Gibbs potential (\ref{Eq: Hydro GibbsFreeEnergy}) can be obtained from (\ref{Eq: CSBH GeneralGibbs}) by a Legendre transform. Before we do this however we need to obtain the charge and particle number from $\tilde{G}$.}

{\ We can obtain the relativistic energy $E$ by computing the ADM mass of (\ref{Eq: CSBH Backgroundmetric}). Using the relativistic equation of state $\epsilon=3P$ and (\ref{Eq: LC NRPressure}) gives us the non-relativistic pressure $P_{nr}$. Finally, using the integrated form of the first law of thermodynamics and the definition of the free energy\footnote{Our formula for the free energy does not appear to agree with the on-shell action of \cite{Adams:2009dm}. An important consistency check of the on-shell action is that one recover the correct entropy i.e.; $S=-\left(T\frac{\partial}{\partial T}+1\right) \frac{\tilde{G}}{T}$. It is not clear to us that the expression in \cite{Adams:2009dm} passes this check. On the contrary, because we are working in the grand canonical ensemble and have derived (\ref{Eq: CSBH FullexpressionGenGibbs}) by calculating the pressure from the ADM mass of the black hole, $\tilde{G}$ definitely passes this test.} implies:
	\begin{eqnarray}
		\tilde{G} &=& E - TS - \mu N - \mu_{q} J^{+} \label{Eq: CSBH GenGibbsLegendreTransform} \\
				  &=& - P_{nr}(T,\mu,\mu_{q}) V_{2} \nonumber \\
				  &=& - \frac{V_{2} \Delta x^{-}}{16 \pi G_{5}} \left(r_{+}^{4} - \frac{8}{3} \frac{\mu_{q}^2}{\mu} r_{+}^2 \right) \label{Eq: CSBH FullexpressionGenGibbs}
	\end{eqnarray}
where $\Delta x^{-}$ has been introduced to characterise the period of the compactified $x^{-}$-direction \cite{Rangamani:2008gi}. Using the thermodynamic relations supplied by:
	\begin{displaymath}
		d \tilde{G} = \underbrace{\left(\frac{\partial G}{\partial V} \right)_{T,\mu,\mu_{q}}}_{-P_{nr}(x)} dV
			 		+ \underbrace{\left(\frac{\partial G}{\partial T} \right)_{V,\mu,\mu_{q}}}_{-S(x)} dT
			 		+ \underbrace{\left(\frac{\partial G}{\partial \mu} \right)_{V,T,\mu_{q}}}_{-N(x)} d\mu
			 		+ \underbrace{\left(\frac{\partial G}{\partial \mu_{q}} \right)_{V,T,\mu}}_{-J^{+}(x)} d\mu_{q}
	\end{displaymath}
we find the following additional quantities:
	\begin{displaymath}
		\begin{array}{ccccc}
			N = - \frac{2P_{nr}V_{2}}{\mu} & \; &
		J^{+} = \frac{\sqrt{3} \beta Q}{2 \pi G_{5}} \Delta x^{-} V_{2} & \; &
			S = \frac{\beta r_{+}^{3}}{4 G_{5}} \Delta x^{-} V_{2}
		\end{array}
	\end{displaymath}
In particular the entropy, at constant $t$, matches that obtained from (\ref{Eq: CSBH HawkingEntropyDensity}). Neither $J^{+}$ nor $\mu_{q}$ attain their relativistic values on setting $\beta=1$ however the combination $\mu_{q} J^{+}$ does. We also note that the energy density calculated by inverting (\ref{Eq: CSBH GenGibbsLegendreTransform}) and dividing by the spatial volume $V_{2}$ satisfies the equation of state for a non-relativistic fluid:
	\begin{eqnarray}
		\epsilon &=& \frac{\Delta x^{-}}{16 \pi G_{5}} \left(r_{+}^{4} - \frac{8}{3} \frac{\mu_{q}^2}{\mu} r_{+}^2 \right) \nonumber \\
		  		 	 &=& P_{nr} \nonumber
	\end{eqnarray}
which acts as a check of our thermodynamics.}

{\ The Gibbs potential is given by the Legendre transform:
	\begin{eqnarray}
		G(P_{nr},T,N,J^{+}) &=& \tilde{G} + P_{nr}V_{2} + \mu N + \mu_{q} J^{+} \nonumber \\
				   	   &=& \mu N + \mu_{q} J^{+} \nonumber
	\end{eqnarray}
where all the quantities on the right-hand side are functions of $P$, $T$, $N$ and $J^{+}$. It is easy to check we get (\ref{Eq: Hydro GibbsFreeEnergyVariation}). Using (\ref{Eq: CSBH FullexpressionGenGibbs}) we find:
	\begin{eqnarray}
		G = - \frac{\Delta x^{-} V_{2} }{8 \pi G_{5}} \left[ r_{+}^4 + \frac{16}{3} \frac{\mu_{q}^2}{\mu} r_{+}^2 \right]
		\label{Eq: CSBH FullexpressionGibbs}
	\end{eqnarray}
The specific heat at constant pressure, particle number and charge is then given by:
	\begin{eqnarray}
		c_{P_{nr}, N, J^{+}} &=& \frac{\pi^2 T}{r_{+}^{2}\left(T, \frac{\mu}{T}, \frac{\mu_{q}}{T}\right)} \left[1 - \frac{4}{3} \frac{\mu_{q}^2}{\mu r_{+}^{2}\left(T, \frac{\mu}{T}, \frac{\mu_{q}}{T}\right)} + \frac{64}{9} \frac{\mu_{q}^{4}}{\mu^2 r_{+}^{4}\left(T, \frac{\mu}{T}, \frac{\mu_{q}}{T}\right)} \right]^{-1}
								 \label{Eq: CSBH SpecificHeat}
	\end{eqnarray}
where $r_{+}\left(T, \frac{\mu}{T}, \frac{\mu_{q}}{T}\right)$ is given by (\ref{Eq: Hydro Defofr}). In the limit that the charge goes to zero this approaches the result for the heat capacity given by \cite{Rangamani:2008gi}. We note that thermodynamic quantities like (\ref{Eq: CSBH FullexpressionGibbs}) and (\ref{Eq: CSBH SpecificHeat}) will also apply to the first-order \Schr fluid of the next section.}

\section{A non-relativistic fluid with Schr\"{o}dinger symmetry} 

{\ In this section we go beyond the \AdSnrCFT as discussed in the previous section and obtain first order corrections to the non-relativistic fluid with \Schr symmetry discussed above. We begin by applying the fluid-gravity procedure of \cite{Bhattacharyya:2008jc} to an asymptotically $\AdS$ charged black hole with translationally invariant horizon as done in \cite{Banerjee:2008th, Erdmenger:2008rm}. We then indicate how the same method could be applied to our asymptotically \Schr black hole metric (\ref{Eq: CSBH 5dmetric}). However we are saved from actually carrying out the latter procedure as we can simply TsT transform the, already calculated, asymptotically $\AdS$, charged black brane with first order fluid corrections. Once this has been accomplished we calculate the boundary stress tensor complex and charge current and identify expressions for the coefficients of (\ref{Eq: Hydro StresstensorI}), (\ref{Eq: Hydro EnergycurrentI}) and (\ref{Eq: Hydro ChargecurrentI}).}

\subsection{Review of the charged, relativistic fluid-gravity correspondence}


{\ We shall begin by reviewing the fluid-gravity correspondence for a relativistic fluid. The reason for doing this is because the procedure for relativistic fluids is more elegant than its non-relativistic counterpart making it easier to become aquainted with the ideas involved. First, we note that there are several solutions to the equations of motion of an Einstein-Maxwell theory with negative cosmological constant that are asymptotically $\AdS$. A solution of interest to us is the ``charged, boosted black brane'' which, written in terms of in-going Eddington-Finkelstein coordinates, has the following form:
	\begin{eqnarray}
		ds_{5}^2 &=& -2u_{\mu} dx^{\mu} dr - r^2 f\left(m^{(0)}, Q^{(0)}, r\right) u_{\mu} u_{\nu} dx^{\mu} dx^{\nu} + r^2 P_{\mu \nu} dx^{\mu} dx^{\nu} \label{Eq: FG PureBlackBrane} \\
	       A_{Q} &=& \frac{\sqrt{3}Q^{(0)}}{2r^2} u_{\mu} dx^{\mu} \label{Eq: FG PureGaugeField}
	\end{eqnarray}
where we have defined:
	\begin{eqnarray}
		f\left(m^{(0)}, Q^{(0)}, r\right) &=& 1 - \frac{m^{(0)}}{r^4} + \frac{\left(Q^{(0)}\right)^2}{r^6} \nonumber \\
		 				 u_{\mu} dx^{\mu} &=& \frac{1}{\sqrt{1-v^2}} \left(- d\tau + v_{i} dx^{i} \right)  \nonumber
	\end{eqnarray}
and introduced $^{(0)}$ to emphasise constant objects. This metric has the nice feature that it is everywhere non-singular except at $r=0$. Note that $\tau$ plays the role of boundary time and as such the boundary metric has the form: $\eta = \mathrm{diag}(-1,1,1,1)$.} 

{\ Solution (\ref{Eq: FG PureBlackBrane}), (\ref{Eq: FG PureGaugeField}) in and of itself is not of interest for this paper as it corresponds to a stationary (modulo Lorentz boosts) fluid. Consider instead promoting the constants $m$, $Q$ and $v^{i}$ to ``slowly-varying'' fields of the boundary coordinates $m(x^{\alpha})$, $Q(x^{\alpha})$ and $v^{i}(x^{\alpha})$. By slowly varying we mean the following; consider expanding these fields about $x=0$. In a local patch about this point whose size is set by the scale of thermodynamic fluctuations from local equilibrium, denoted $L$, we can use coordinate redefinitions and velocity field gauge choices to set $m(0)=m^{(0)}$, $v^{i}(0)=0$ and $Q(0)=Q^{(0)}$. It can then be seen that higher derivative corrections to $m(x^{\alpha})$, $Q(x^{\alpha})$ and $v_{i}(x^{\alpha})$ are suppressed by higher powers of $\frac{1}{TL}$ where $T(x^{\alpha})$ is the local temperature of the fluid. We should also note that the non-zero temperature and velocities have broken the $SO(4,2)$ symmetry in the boundary theory and thus we can loosely identify these fields as Goldstone bosons\footnote{The fluid stress tensor will be expanded in terms of these Goldstone fields so it can be partitioned into collections of terms with the same factors of $\frac{1}{TL}$.}.}

{\ The line element and gauge field with promoted constants has the form:
	\begin{eqnarray}
		ds_{5}^2  &=& -2u_{\mu}(x^{\alpha}) dx^{\mu} dr - r^2 f(m(x^{\alpha}),Q(x^{\alpha}),r) u_{\mu}(x^{\alpha}) u_{\nu}(x^{\alpha}) dx^{\mu} dx^{\nu} 
									+\nonumber \\
						&\;& + r^2 P_{\mu \nu}(x^{\alpha}) dx^{\mu} dx^{\nu} \label{Eq: FG ZerothOrderFluidMetric} \\
			A_{Q} &=& \frac{\sqrt{3} Q(x^{\alpha})}{2r^2} u_{\mu}(x^{\alpha}) dx^{\mu} \label{Eq: FG ZerothOrderFluidGauge}
	\end{eqnarray}
Generically the corresponding metric does not satisfy our equations of motion. However, in \cite{Banerjee:2008th} terms of higher order in $\frac{1}{TL}$ were added to the above ansatz and treated as perturbations upon it. It was discovered that if the $m(x^{\alpha})$, $Q(x^{\alpha})$ and $v^{i}(x^{\alpha})$ fields additionally obey the following constraints:
	\begin{eqnarray}
		\left. \partial_{\tau} m(x^{\alpha}) \right|_{x=0} &=& - \frac{4}{3} m(x^{\alpha}) \left. \partial_{i} v^{i}(x^{\alpha}) \right|_{x=0} \label{Eq: FGC ConservationV} \\
		   \left. \partial_{i} m(x^{\alpha}) \right|_{x=0} &=& - 4 m(x^{\alpha}) \left. \partial_{\tau} v_{i}(x^{\alpha}) \right|_{x=0} \label{Eq: FGC ConservationI} \\
		\left. \partial_{\tau} Q(x^{\alpha}) \right|_{x=0} &=& - Q(x^{\alpha}) \left. \partial_{i} v^{i}(x^{\alpha}) \right|_{x=0} \label{Eq: FGC ConservationQ}
	\end{eqnarray}
then the Einstein-Maxwell equations, given by variation of (\ref{Eq: CSBH TypeIIBRelReduction}), are satisfied at first order. This procedure can be extended to higher orders but our concern here lies only with first order corrections.}

{\ To detail this process of solving the Einstein-Maxwell equations more carefully we note that the metric (\ref{Eq: FG ZerothOrderFluidMetric}) and gauge field (\ref{Eq: FG ZerothOrderFluidGauge}) can be expanded about the point $x^{\mu}=0$ to give:
	\begin{eqnarray}
		ds_{5}^{2} &=& 2d\tau dr - r^2 f\left(m^{(0)}, Q^{(0)}, r \right) \left(d \tau\right)^2 + r^2 dx_{i} dx^{i} - \nonumber \\
			      &\;& - 2 x^{\mu} \partial_{\mu} v^{(0)}_{i} dx^{i} dr - 2 x^{\mu} \partial_{\mu} v^{(0)}_{i} r^2 \left[ 1-f\left(m^{(0)}, Q^{(0)}, r \right) \right] dx^{i} d\tau + \nonumber \\
			      &\;& + \left[ \frac{x^{\mu} \partial_{\mu} m^{(0)}}{r^2} - \frac{2Q^{(0)} x^{\mu} \partial_{\mu} Q^{(0)} }{r^4} \right] \left(d\tau\right)^2 \label{Eq: FG Expandedfluidmetric} \\
		    A_{Q}  &=& - \frac{\sqrt{3}}{2} \left[ \left( \frac{Q^{(0)} + x^{\mu} \partial_{\mu} Q^{(0)} }{r^2} \right) d\tau - \frac{Q^{(0)}}{r^2} x^{\mu} \partial_{\mu} v^{(0)}_{i} dx^{i} \right] 
						\label{Eq: FG Expandedfluidgauge}
	\end{eqnarray}
where as previously mentioned we have used coordinate scalings and a suitable gauge choice for the velocity field to set:
	\begin{eqnarray} 
		m(x^{\mu}) &=& m^{(0)} + x^{\mu}\partial_{\mu} m^{(0)} + \ldots  \nonumber \\
		Q(x^{\mu}) &=& Q^{(0)} + x^{\mu}\partial_{\mu} Q^{(0)} + \ldots  \nonumber \\
		u(x^{\mu}) &=& -d\tau + x^{\mu} \partial_{\mu} v^{(0)}_{i} dx^{i} + \ldots \nonumber
	\end{eqnarray}
Note that ultra-locality of our field equations allows us to extend our solution away from $x^{\mu}=0$ to the entire manifold \cite{Bhattacharyya:2008jc}. Our metric (\ref{Eq: FG PureBlackBrane}) and gauge field (\ref{Eq: FG PureGaugeField}) have spatial $SO(3)$ symmetry and therefore we can parameterise the corrections to (\ref{Eq: FG ZerothOrderFluidMetric}) and (\ref{Eq: FG ZerothOrderFluidGauge}) in the following form:
	\begin{eqnarray}
			g^{(1)} &=& -3 h^{(1)}(r) d\tau dr + \frac{k^{(1)}(r)}{r^2} \left(d\tau\right)^2 + 3 r^2 h^{(1)}(r) \left(dx^{i}\right)^2	+ \nonumber \\
						 &\;& + 2 r^2 \left[1-f\left(m^{(0)}, Q^{(0)}, r \right)\right] j^{(1)}_{i}(r) d\tau dx^{i} 
									+ r^2 \alpha^{(1)}_{ij}(r) dx^{i} dx^{j} \label{Eq: FC MetricCorrection} \\
		A^{(1)}_{Q} &=& \frac{ \sqrt{3} w^{(1)}(r) }{2r^2} d\tau - \left[ \frac{\sqrt{3} Q^{(0)} }{2r^2} j^{(1)}_{i}(r) - g_{i}^{(1)}(r) \right] dx^{i} \label{Eq: FC GaugeCorrection}
	\end{eqnarray}
where the first three terms of (\ref{Eq: FC MetricCorrection}) are scalar, the fourth term vector and the fifth term a symmetric 2-tensor with respect to $SO(3)$. Similarly the first term in $A^{(1)}_{Q}$ is a scalar and the second term a vector. If we denote the Einstein equations by $E_{ab}$ and the Maxwell equations $M_{a}$ then when $E_{ab}$ and $M_{a}$ are evaluated on the metric and gauge fields with first order corrections they can also be decomposed with respect to $SO(3)$. We can further separate the Einstein-Maxwell equations in each of these sectors into two groups; constraint equations\footnote{To call the projected equations constraints is a slight abuse of terminology since we are not dealing with an initial value problem here.}, which are obtained by contracting the tensors $E_{ab}$ and $M_{a}$ with the vector dual to the one-form $(dr)^{a}$, and dynamical equations. The constraint equations depend on the particular nature of the fluid we are considering in that, as they correspond to covariant conservation of SEM tensor and charge currents (\ref{Eq: FGC ConservationV})-(\ref{Eq: FGC ConservationQ}), we need to solve the hydrodynamic equations for temperature and charge profiles when given a velocity distribution. However, the dynamical equations can be completely solved by choosing the correct dependence of the undetermined functions in (\ref{Eq: FC MetricCorrection}) and (\ref{Eq: FC GaugeCorrection}) on $m$, $Q$ and $r$.}

{\ A relativistic and gauge independent expression for the charged black brane with $\AdS[5]$ boundary found in \cite{Banerjee:2008th} to first order in derivatives is:
	\begin{eqnarray}
		ds_{5}^{2} &=& -2u_{\mu} dx^{\mu} dr - r^2 f(m,Q,r) u_{\mu} u_{\nu} dx^{\mu} dx^{\nu} + r^2 P_{\mu \nu} dx^{\mu} dx^{\nu} - \nonumber \\
				  		&\;& - 2 r u_{\mu} \left(u^{\lambda} \nabla_{\lambda} u_{\nu}\right) dx^{\mu} dx^{\nu} 
				  		 		 + \frac{2}{3} r \left( \nabla_{\lambda} u^{\lambda} \right) u_{\mu}  u_{\nu} dx^{\mu} dx^{\nu} + \nonumber \\
							&\;& + 2 \frac{r^2}{r_{+}} F_{1}(m,Q,r) \sigma_{\mu \nu} dx^{\mu} dx^{\nu} 
				   		 		 - \frac{2 \sqrt{3} \kappa_{CS} Q^3}{m r^4} u_{\mu} l_{\nu} dx^{\mu} dx^{\nu} - \nonumber \\
				   		&\;& - 12 Q \frac{r^2}{r_{+}^7} F_{2}(m,Q,r) u_{\mu} \left( 		P^{\lambda}_{\nu} \nabla_{\lambda} 
					   			 + 3 u^{\lambda} \nabla_{\lambda} u_{\nu} \right) Q dx^{\mu} dx^{\nu} \label{Eq: FOF OriginalMetric} \\
			A_{Q}  &=& \frac{\sqrt{3} Q}{2r^2} u_{\mu} dx^{\mu} + \nonumber \\
				 &\;& + \frac{3 \kappa_{CS} Q^{2}}{m r^2} l_{\mu} dx^{\mu} + \frac{ \sqrt{3} r^{5} }{2 r_{+}^7 } \left[\frac{\partial}{\partial r} F_{2}(m,Q,r)\right] \left( P^{\lambda}_{\mu} \nabla_{\lambda} 
					  + 3 u^{\lambda} \nabla_{\lambda} u_{\mu} \right) Q dx^{\mu} \label{Eq: FOF Gaugefield}
	\end{eqnarray}
where the first line of each definition is zeroth order in fluid derivatives, the subsequent lines first order and:
	\begin{eqnarray}
		f(m,Q,r_{+}) &=& 0 \nonumber \\
		F_{1}(m,Q,r) &=& \int_{\frac{r}{r_{+}}}^{\infty} dx \frac{x \left(x^2 + x +1 \right)}{\left(x+1\right)\left(x^{4}+x^{2}-\frac{Q^2}{r_{+}^{6}}\right)}   \nonumber \\
		F_{2}(m,Q,r) &=& \frac{1}{3} \left(1-\frac{m}{r^{4}}+\frac{Q^{2}}{r^{6}}\right) \int_{\frac{r}{r_{+}}}^{\infty} dx \frac{1}{\left(1-\frac{m}{x^{4}}+\frac{Q^{2}}{x^{6}}\right)^2} 
						 \left(\frac{1}{x^{8}} - \frac{3}{4 x^{7}} \left(1 + \frac{r_{+}^{4}}{m} \right) \right) \nonumber
	\end{eqnarray}
These results were obtained by perturbatively solving the equations of motion from (\ref{Eq: CSBH TypeIIBRelReduction}) with trivial dilaton field using the fluid derivative expansion method of \cite{Bhattacharyya:2008jc}. It has been shown that this solution can readily be uplifted to a 10-dimensional solution of Type IIB by adding a suitably deformed $S^{5}$ term to the metric and folding an $A_{Q}$ term into the standard expression for the Ramond-Ramond five form \cite{Buchel:2006gb}.}

\subsection{Charged non-relativistic fluids}

{\ With the well-established relativistic fluid-gravity derivative expansion as a skeleton we can now consider moving beyond zeroth order for a fluid with \Schr symmetry. In particular we shall find the derivative expansion corrections to (\ref{Eq: CSBH 5dmetric}) at first order. We begin by noting that (\ref{Eq: CSBH 5dmetric}) is not regular across the future horizon. One of the nice properties of (\ref{Eq: FG ZerothOrderFluidMetric}) was its regularity at all points other than $r=0$. This can be remedied by translating the $x^{+}$ and $x^{-}$ coordinates in the following manner:
	\begin{eqnarray}
		dx^{+} &\rightarrow& dx^{+} - \frac{\beta}{r^2 f(m,Q,r)} dr \nonumber \\
		dx^{-} &\rightarrow& dx^{-} + \frac{1}{2 \beta r^2 f(m,Q,r)} dr \nonumber
	\end{eqnarray}
Our metric, gauge field and massive vector field correspondingly become:
	\begin{eqnarray}
		(ds_{5}^2)'' &=& \frac{r^2}{k} \left[ \left(\frac{1-f(m,Q,r)}{4\beta^2} - r^2 f\right) \left( dx^{+} \right)^{2} + \beta^2 \left(1-f(m,Q,r) \right) \left( dx^{-} \right)^{2} + \right. \nonumber \\ 
								&\;& \phantom{\frac{r^2}{k} \left[ \right.} + (1+f(m,Q,r)) dx^{+} dx^{-} +\left(\frac{1}{\beta r^2} + 2 \beta\right) dx^{+} dr - \left(\frac{2 \beta}{r^2}\right) dx^{-} dr +  \nonumber \\
								&\;& \left. \phantom{\frac{r^2}{k} \left[ \right.} + k d\vec{x}^2 \right] - \beta^2 \frac{dr^2}{k} \label{Eq: CSBH 5dmetricRegular} \\
			   A_{Q} &=& \frac{\sqrt{3}}{2} \frac{Q}{r^2} \left[ \frac{dx^{+}}{2\beta} - \beta dx^{-} \right] \\
			   A_{M} &=& \frac{\beta r^2}{k} \left[ \left(1+ f(m,Q,r)\right) \frac{dx^{+}}{2\beta} + \left(1 - f(m,Q,r) \right) \beta dx^{-} - \frac{dr}{r^2} \right] \label{Eq: CSBH MassiveVectorRegular}
	\end{eqnarray}
where we have used a gauge choice to remove a $dr$ term from $A_{Q}$. It should be noted that because of the dominant scaling of the $x^{+}$ term in (\ref{Eq: CSBH 5dmetricRegular}) we lack a well-defined boundary metric for the asymptotically \Schrcomma charged black hole.}

{\ Our boundary theory has Galilean symmetry and by boosting (\ref{Eq: CSBH 5dmetricRegular}) we obtain a class of solutions with the same thermodynamics but non-zero velocity. The boosted solutions will be classified by five constants $\beta$, $v^{i}$, $Q$ and $m$ however these contain no new physics. Instead, to obtain results other than the ideal fluid, we need to promote the parameters $m$, $Q$, $\beta$ and $v^{i}$ to functions of the boundary coordinates and find first order fluid corrections to our fields. A local Galilean boost has the form:
	\begin{eqnarray}
		\vec{x} &\rightarrow& \vec{x} + \vec{v}(x) x^{+} \nonumber \\
		  x^{-} &\rightarrow& x^{-} + \vec{v}(x) \cdot \vec{x} + \frac{1}{2} \vec{v}^2(x) x^{+} \nonumber
	\end{eqnarray}
Consider the vicinity of a point $x^{\mu}=0$ and use global Galilean invariance to set the velocity at this point to zero. Performing our local boost on (\ref{Eq: CSBH 5dmetricRegular})-(\ref{Eq: CSBH MassiveVectorRegular}), to one derivative in velocity, we find the following additional terms:
	\begin{eqnarray}
			g^{(1)} &=& \frac{2 \beta^2 r^2}{k} \left[ (1-f(m,Q,r)) dx^{-} \vec{x}  + \frac{1}{2}(1+f(m,Q,r)) dx^{+} \vec{x}  + \frac{1}{\beta r^2} \vec{x} dr + \frac{k x^{+}}{\beta^2} d\vec{x} \right] \cdot d \vec{v} \nonumber \\
		A^{(1)}_{Q} &=& - \frac{\sqrt{3}}{2} \frac{Q}{r^2} \beta \vec{x} \cdot d \vec{v} \nonumber \\
		A^{(1)}_{M} &=& + \frac{\beta^2 r^2}{k} (1-f(m,Q,r)) \vec{x} \cdot d \vec{v} \nonumber
	\end{eqnarray}
Again, generically, the metric, gauge field and massive vector field with the above terms will not satisfy the equations of motion from (\ref{Eq: CSBH SchrAction}) and as such we need to find suitable corrections. Our precursor fields (\ref{Eq: CSBH 5dmetric})-(\ref{Eq: CSBH MassiveGauge}) have $SO(2)$ spatial invariance and thus we can parameterise our corrections with respect to this symmetry much as we did with the $SO(3)$ symmetry in the relativistic case (\ref{Eq: FC MetricCorrection}), (\ref{Eq: FC GaugeCorrection}). However we face two additional complications for the asymptotically \Schr spacetime. Firstly, we not only have new equations of motion for the massive vector field and dilaton to satisfy but our equations of motion have become more complex. Secondly, the $SO(2)$ symmetry is not as helpful in the \Schr case as the $SO(3)$ was in the relativistic case. For example, in the \Schr case there are two possible vector sectors in the metric coming from $dx^{+} dx^{i}$ and $dx^{-} dx^{i}$ compared to one, $d\tau dx^{i}$, for a relativistic fluid.}

{\ It is clear then that solving the equations of motion from (\ref{Eq: CSBH SchrAction}) to first order in derivatives would be a cumbersome process. Instead, as previously mentioned, we can perform a TsT transformation of (\ref{Eq: FOF OriginalMetric}) and (\ref{Eq: FOF Gaugefield}) to obtain the $U(1)$ charged, \Schr fluid at first order. As we previously assumed trivial $x^{-}$ dependence in our hydrodynamic variables in order to light-cone reduce, we can drop all $x^{-}$ dependence in the metric coefficients. This ties in nicely with the fact that to use the TsT solution generating technique detailed in the appendix we require $x^{-}$ and $\psi$ to be isometry directions of the metric. In particular notice that $x^{-}$ is a null isometry direction in the boundary theory and so the TsT with a twist along this direction will coincide with an NMT of our fields.} 

{\ Using the identities from \cite{Rangamani:2008gi}, in particular the zeroth order current conservation equations, our results indicate that the dilaton now has the form:
	\begin{eqnarray}
		e^{-2\Phi''} &=& k \nonumber \\
				   	 &=& 1 + \beta^2 r^2 \left[ 1 - f(m,Q,r) \right] + \frac{2 \sqrt{3} \beta^3 Q^{3} }{m r^4} \kappa_{CS} \epsilon^{ij} \partial_{i} v_{j}
	\end{eqnarray}
where $i,j$ run over the spatial directions $\left\{x,z\right\}$ and we have taken $u^{+}=\beta$. The full metric at first order is given by:
	\begin{eqnarray}
		(ds_{5}^2)'' &=& -2u_{\mu} dx^{\mu} dr - r^2 f(m,Q,r) u_{\mu} u_{\nu} dx^{\mu} dx^{\nu} + r^2 P_{\mu \nu} dx^{\mu} dx^{\nu} - \nonumber \\
				    &\;& -2 r u_{\mu} \left(u^{\lambda} \nabla_{\lambda} u_{\nu}\right) dx^{\mu} dx^{\nu} + \frac{2}{3} r \left( \nabla_{\lambda} u^{\lambda} \right) u_{\mu}  u_{\nu} dx^{\mu} dx^{\nu} 
						 + 2 \frac{r^2}{r_{+}} F_{1}(m,Q,r) \sigma_{\mu \nu} dx^{\mu} dx^{\nu} - \nonumber \\
			  	    &\;& - \frac{2 \sqrt{3} \kappa_{CS} Q^3}{m r^4} u_{\mu} l_{\nu} dx^{\mu} dx^{\nu} - 12 Q \frac{r^2}{r_{+}^7} F_{2}(m,Q,r) u_{\mu} \left( P^{\lambda}_{\nu} \nabla_{\lambda} 
					     + 3 u^{\lambda} \nabla_{\lambda} u_{\nu} \right) Q dx^{\mu} dx^{\nu} - \nonumber \\
				    &\;& - k \left(A_{M}\right)_{\mu} \left(A_{M}\right)_{\nu} dx^{\mu} dx^{\nu} \label{Eq: FOF FOFSchrMetric}
	\end{eqnarray}
The massive vector field, in the light-cone coordinate system, has the form:
	\begin{eqnarray}
    	A_{M} &=& \frac{1}{k}  \left[ \left( \beta^2 r^2 \left(1-f(m,Q,r) \right) + \frac{2 \sqrt{3}}{mr^4} \kappa_{CS} Q^{3} \beta^3 \epsilon_{jk} \partial^{j} v^{k} \right) dx^{-} + \left(r^2 dx^{+} - \beta dr \right) 			  				+ \right. \nonumber \\
    		 &\;& \phantom{\frac{1}{k}  \left[ \right.} + \left( \beta r^2 \left(1-f(m,Q,r) \right) + \frac{\sqrt{3} \kappa_{CS} Q^3 \beta^2}{mr^4} \epsilon_{jk} \partial^{j} v^{k} \right) u_{\alpha} dx^{\alpha} + \nonumber \\
			 &\;& \phantom{\frac{1}{k}  \left[ \right.} + \left( 2\frac{r^2}{r_{+}} F_{1} \sigma_{-\alpha} - \frac{\sqrt{3} \kappa_{CS} Q^3 \beta}{mr^4} l_{\alpha} - 6 \beta Q \frac{r^2}{r_{+}^{7}} F_{2} \left( \nabla_{\alpha} Q -\frac{3}{2} Q \frac{\nabla_{\alpha} P_{nr}}{\epsilon_{nr} + P_{nr}} \right) \right. \nonumber \\
			 &\;& \left. \left. \phantom{\frac{1}{k}  \left[ \right. \left( \right.} + \frac{\beta r}{2} \frac{\nabla_{\alpha} P_{nr}}{\epsilon_{nr} + P_{nr}} \right) dx^{\alpha} \right] \nonumber
	\end{eqnarray}
where $\alpha \in \left\{+,x,z\right\}$ and $A_{Q}$ is unchanged from (\ref{Eq: FOF Gaugefield}) by the TsT. Note that all but the last line of the metric (\ref{Eq: FOF FOFSchrMetric}) occurs in (\ref{Eq: FOF OriginalMetric}) so that all the deformation comes from the vector field $A_{M}$. This makes sense in light of the fact that the massive vector field is the only additional dynamical field between the Melvinised and un-Melvinised solutions.}

\subsection{Transport coefficients of non-relativistic fluids}

{\ Now that the bulk metric and vector fields are known we can consider the fluid side of the fluid-gravity correspondence. To calculate the hydrodynamic and charge coefficients we need to determine the boundary value of the conserved currents associated with the bulk metric and the gauge field. However, we face a particular problem in \Schr spacetimes due to the slow asymptotic fall-off of the modes. We follow \cite{Maldacena:2008wh}, \cite{Ross:2009ar} and \cite{Rangamani:2008gi} and interpret the SEM tensor of the asymptotically $\AdS$ theory prior to the TsT transformation as a tensor complex (collection of fields) in the non-relativistic theory. This complex is given exactly by the identifications in (\ref{Eq: LC TensorComplexIdentification}).}

{\ The boundary SEM tensor and charge current for the $\RNAdS[5]$ black hole have the forms:
	\begin{eqnarray}
		T_{\mu \nu} &=& P \left( \eta_{\mu \nu} + 4 u_{\mu} u_{\nu} \right) - 2 \eta \sigma_{\mu \nu} \nonumber \\
		j^{\mu}     &=& q u^{\mu} - \kappa_{q} \left( P^{\mu \nu} \nabla_{\nu} q + 3 q u^{\nu} \nabla_{\nu} u^{\mu}  \right) - \mho l^{\mu} \label{Eq: FG Boundarycurrent}
	\end{eqnarray}
The coefficients, in terms of $G_{5}$ which can be related to the central charge of the field theory, are:
	\begin{displaymath}
		\begin{array}{ccccccc}
		      \epsilon = 3 P & \; &
		 	         P = \frac{m}{ 16 \pi G_{5} } & \; &
		          \eta = \frac{ r_{+}^{3} }{ 16 \pi G_{5} } & \; & \\
			         q = \frac{ \sqrt{3} Q }{ 4 \pi G_{5} } & \; &
		    \kappa_{q} = \left(\frac{r_{+}^{4} + m}{4 m r_{+}} \right) & \; &
		        \gamma = - \frac{3q}{4\epsilon} \kappa_{q} & \; &
		   	      \mho = - \frac{\kappa_{CS}}{2P} q^2
		\end{array}
	\end{displaymath}
where we have applied the SEM conservation equation at zeroth order in derivatives:
	\begin{displaymath}
		u^{\nu} \nabla_{\nu} u^{\mu} + \frac{P^{\mu \nu} \nabla_{\nu} P}{\epsilon + P} = 0
	\end{displaymath}
to the penultimate term of (\ref{Eq: FG Boundarycurrent}) to write it in terms of $\epsilon$. Note that our parity violation coefficient $\mho$ is indeed determined by the Chern-Simon's parameter $\kappa_{CS}$, the charge $q$ and the fluid pressure $P$.}

{\ To convert our relativistic results into their non-relativistic counterparts we need to fix the normalisation of our boundary velocity $u^{\mu}$. Fortunately we have already isolated a suitable choice in (\ref{Eq: CSBH NullKilling}) when we fixed the coefficient of $(\partial_{+})^{a}$ in the horizon null generator, $\xi^{a}$, to be unit. Hence we take:
	\begin{eqnarray}
		u^{+} &=& \beta \nonumber \\
		u^{i} &=& \beta v^{i} - \frac{\beta \eta}{\rho} \delta^{ij} \left( \partial_{j} \beta - \frac{\beta}{\left(\epsilon+P\right)} \partial_{j} P  \right) \label{Eq: FOF Vi}
	\end{eqnarray}
Using the maps (\ref{Eq: LC NRStressTensor})-(\ref{Eq: LC Energycurrent}), (\ref{Eq: LC TemporalPart}) and (\ref{Eq: LC SpatialPart}) it is now possible to determine all the non-relativistic quantities in terms of $m$, $Q$, $\beta$ and $v^{i}$. Modulo the subtlety involving the thermal conductivity which we shall discuss next the zeroth order coefficients are:
	\begin{eqnarray}
		\begin{array}{ccccc}
			\epsilon_{nr} = P_{nr} & \; &
		 	 	   P_{nr} = \frac{ m }{ 16 \pi G_{5} } \Delta x^{-} & \; &
					 \rho = - \frac{ 2 P_{nr} }{ \mu }
		\end{array} \nonumber \\
		q_{nr} = \frac{ \sqrt{3} \beta Q }{	4 \pi G_{5} } \left[ 1 + \frac{\sqrt{3} \beta Q \kappa_{CS}}{m} \epsilon_{jk} \partial^{j} v^{k} \right] \Delta x^{-} 
		\label{Eq: FOF NRBoundaryQuantities}
	\end{eqnarray}
The apparent disparity for $q_{nr}$ with the thermodynamic result is due to the previous decomposition of the combination $\mu_{q} J^{+}$ into $\mu_{q}$ and $J^{+}$ which shuffled a factor of two between them. At first order we also have:
	\begin{eqnarray}
		\begin{array}{ccccc}
			  \eta_{nr} = \frac{\beta r_{+}^{3}}{16 \pi G_{5}} \Delta x^{-} & \; &
			\kappa_{nr} = \left(\frac{r_{+}^{4} + m}{4 m \beta r_{+}} \right) \Delta x^{-} & \; &
			  \mho_{nr} = - \frac{\kappa_{CS}}{2 P_{nr}} q_{nr}^2
		\end{array} \label{Eq: FOF NRViscosity} \\
		  \gamma^{ij}_{nr} = \left(\frac{1}{4 \epsilon_{nr}} \right) 
												   \left[ q_{nr} \left(3 \frac{\eta_{nr}}{\rho} - \kappa_{nr} \right) \delta^{ij} 
														      + 4 \frac{\mho_{nr} \epsilon_{nr}}{\rho} \epsilon^{ij} \right] \Delta x^{-} \nonumber \\
		\digamma^{ij}_{nr} = - \left( \frac{1}{2 \rho} \right) \left[ q_{nr} \left(\kappa_{nr}+\frac{\eta_{nr}}{\rho}\right) \delta^{ij} 
													 + 4 \frac{\mho_{nr} \epsilon_{nr}}{\rho} \epsilon^{ij} \right] \Delta x^{-} \nonumber
	\end{eqnarray}
where the $\Delta x^{-}$ factors were introduced to ensure the above quantities are volume densities with respect to the two-dimensional spatial volume $V_{2}$. The parity violating term $\mho_{nr}$ already multiplies an object that is first order in fluid derivatives, see (\ref{Eq: LC SpatialPart}). Hence when expressing it in terms of $q_{nr}$ we have dropped any additional velocity derivatives. Similarly for replacing $\gamma$ in $\gamma^{ij}_{nr}$.}

{\ We would now like to extract the thermal conductivity. Re-expressing the final two terms of (\ref{Eq: LC Energycurrent}) as the differential of a logarithm and using (\ref{Eq: CSBH HawkingTemp}) and (\ref{Eq: FOF NRBoundaryQuantities}) we can write:
	\begin{eqnarray}
			- \frac{2 \eta_{nr} P_{nr}}{\rho} \delta^{ij} \partial_{j} \ln{ \left(\frac{P_{nr}^{\frac{3}{2}}}{\rho} \right) }
		&=& - \frac{4 \eta_{nr} P_{nr}}{\rho T} \delta^{ij} \partial_{j} T - \nonumber \\
		&\;& 	- \frac{2 \eta_{nr}}{\rho} \delta^{ij} \partial_{j} \left[ \frac{1}{2} \ln \left(1+\frac{Q^2}{r_{+}^{6}}\right) - 2 \ln\left(1-\frac{Q^2}{2 r_{+}^{6}}\right)  \right] \label{Eq: FOF ThermalConduct}
	\end{eqnarray}
If we interpret the thermal conductivity as the coefficient of the term with no explicit charge dependence then $\kappa_{T}$ has the same functional dependence on $\eta_{nr}$, $P_{nr}$, $\rho$ and $T$ as in the uncharged case of \cite{Rangamani:2008gi}:
	\begin{equation}
		\kappa_{T} = \frac{4 \eta_{nr} P_{nr}}{\rho T}
		\label{Eq: FOF ThermalConductivity}
	\end{equation}
As promised there is a new term which vanishes if the local charge density is set to zero. On substituting for $Q$ in terms of $\mu_{q}$ using (\ref{Eq: CSBH ChemicalPotential}) and rearranging we obtain the final term of (\ref{Eq: Hydro EnergycurrentI}) with $\varpi=\kappa_{T}$.}

{\ We would now like to calculate the Prandtl number for the fluid. We first note that the kinematic viscosity, which compares the importance of viscous to inertial forces, is defined by:
	\begin{displaymath}
		\nu = \frac{\eta_{nr}}{\rho}
	\end{displaymath}
where $\rho$ is representative of inertia. The thermal diffusivity is defined by:
	\begin{displaymath}
		\chi = \frac{\kappa_{T}}{\rho c_{P_{nr},N,J^{+}}}	
	\end{displaymath}
where $\kappa_{T}$ measures heat flow from a region of local equilibrium while $\rho c_{P_{nr},N,J^{+}}$ measures the ability of the region to adjust its temperature to match its surroundings. Thus when $\chi$ is large the region in question quickly responds to the temperature of neighbouring regions and equilibriates its temperature.}

\FIGURE{
	\centering
	\includegraphics[width=10cm]{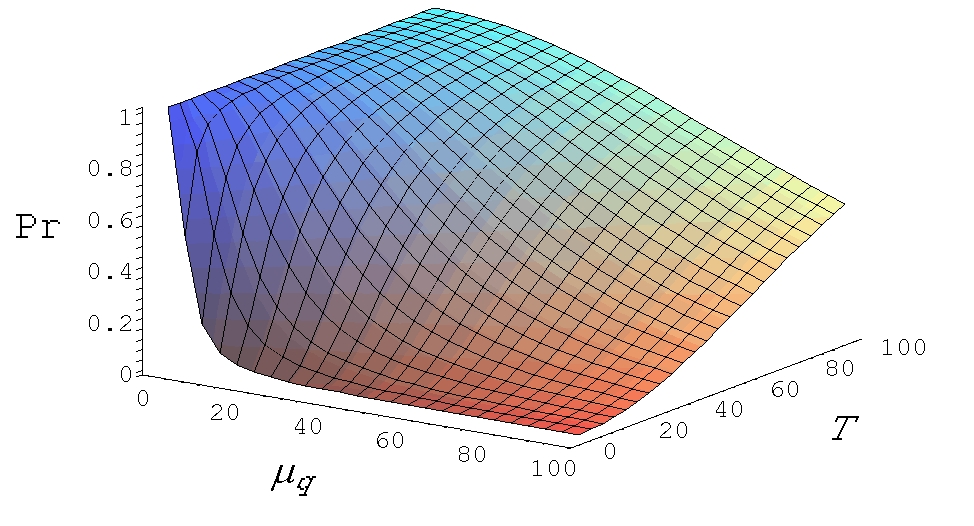}
	\caption{A graph of the Prandtl number against $\mu_{q}$ and $T$. We note that the Prandtl number clearly tends to one when the charge vanishes.}
	\label{Fig: FOF Prandtlnumber}
}

{\ The Prandtl number is given by the ratio of kinematic viscosity to thermal diffusivity:
	\begin{displaymath}
		\Prt = \frac{\nu}{\chi}
	\end{displaymath}
and thus represents the relative importance of viscous effects and heat conduction in reaching steady state flow. Using (\ref{Eq: CSBH SpecificHeat}), (\ref{Eq: FOF NRBoundaryQuantities}), (\ref{Eq: FOF NRViscosity}) and (\ref{Eq: FOF ThermalConductivity}) we find this number to be:
	\begin{displaymath}
		\Prt = \frac{\pi^2 T^2}{2}\left[\frac{4}{3} \mu_{q}^2 - \mu r_{+}^2\left(T, \frac{\mu}{T}, \frac{\mu_{q}}{T} \right) - \frac{64}{9} \frac{\mu_{q}^4}{\mu r_{+}^2\left(T, \frac{\mu}{T}, \frac{\mu_{q}}{T} \right)} \right]^{-1}
	\end{displaymath}
where $r_{+}\left(T, \frac{\mu}{T}, \frac{\mu_{q}}{T} \right)$ is given by (\ref{Eq: Hydro Defofr}). Disappointingly this indicates that the fluid does not achieve a universal value in the presence of a conserved electric charge unlike the uncharged case of \cite{Rangamani:2008gi} where it is identically one. Of note is the fact that $\Prt$ is independent of the particle number chemical potential and compactification radius $\Delta x^{-}$. Figure (\ref{Fig: FOF Prandtlnumber}) is a diagram giving an indication of the dependence of $\Prt$ on charge chemical potential and temperature.}

\section{Discussion}

{\ We began our study by laying the foundations for non-relativistic hydrodynamics in reasonably general terms. We then demonstrated how to obtain the hydrodynamic derivative expansions of a fluid with \Schr invariance at first order from a parent charged, conformal, relativistic theory. Specifically we have shown how to generalise the maps of \cite{Rangamani:2008gi} to the case of an $U(1)$ charged fluid at first order.}

{\ We then focused upon the hydrodynamic limit of a particular three dimensional, non-relativistic conformal field theory and its dual solution which is an asymptotically $\Sch[5]$, charged black hole. Using the TsT technique on an $\RNAdS[5]$ precursor we constructed an action whose equations of motion had the desired black hole as a solution and isolated the thermodynamics. We then obtained first order corrections to the corresponding black brane by using a method based on the derivative procedure detailed in \cite{Bhattacharyya:2008jc}. Because of its relative simplicity we first showed how to find corrections to our initial $\RNAdS[5]$ ansatz before discussing how the same method could be applied in the asymptotically \Schr case. Although in principle we could then have computed the first order corrections to our asymptotically \Schrcomma charged black hole using the fluid-derivative procedure we noted that it would be particularly cumbersome to do so and hence opted instead to use the TsT technique. Thus we arrived at expressions for the metric, gauge field, massive vector field and dilaton to first order in derivatives.}

{\ With the corrected fields to hand in principle it was possible to calculate the asymptotic values of the metric and gauge field directly to determine their corresponding conserved boundary currents. While this is simple to do for the gauge field ambiguities in asymptotic fall off of the metric necessitated that we interpret the boundary SEM tensor in the precursor asymptotically \AdS theory as a tensor complex of the \Schr invariant theory \cite{Maldacena:2008wh}. With these identifications it was relatively simple to apply the \AdSnrCFT to compute the boundary coefficients and with a little work obtain the Prandtl number. An important result discovered here was that the universal value of one for the Prandtl number of an uncharged fluid no longer holds when there is an additional non-zero charge. This suggests that it may be interesting to understand the consequences of a scaling where the non-relativistic charge and particle number were related as this would naturally be interpreted as the charge being carried by the fluid particles. We leave this for future work.}

{\ Although our study concentrated on a fluid occupying two spatial dimensions, the derivative expansions (\ref{Eq: Hydro StresstensorI}), (\ref{Eq: Hydro EnergycurrentI}) and (\ref{Eq: Hydro ChargecurrentI}) apply in any dimension with the caveat that the relativistic one-derivative parity violating term only exists in four dimensions. Similarly the generalisations to multiple $U(1)$ charges or indeed different internal symmetries seems clear and we can determine the hydrodynamic coefficients if we can find a suitable dual black hole with the required asymptotics. A more significant limitation of this paper is reflected in the fact that our fluid is both conformal and incompressible. To remedy this situation we would need to show that we can decouple density fluctuations. As was mentioned previously under a suitable scaling it is possible to reproduce the incompressible Navier-Stokes equations \cite{Bhattacharyya:2008kq} by suppressing sound modes. However this limit has an entirely different symmetry group and the charged and uncharged solutions are trivially related.}

\paragraph{Acknowledgements:}

{\ Thanks go to Mukund Rangamani for guidance in producing this paper and also to Simon Ross and Paolo Benincasa for several valuable discussions. I would like to thank Derek Harland for his mathematical assistance. Finally, I would like to acknowledge the STFC for financial support.}

\appendix

\section{Mathematical conventions}

\subsection{Hodge duals}

{\ Following \cite{Adams:2009dm} we note that the ten-dimensional Hodge dual on $(\RNAdS[5]) \times S^{5}$ can be restricted to the 5-dimensional $\RNAdS[5]$ manifold, 1 dimensional fibration coordinate and $\mathbb{CP}^{2}$ in the following manner:
	\begin{displaymath}
		*_{10} = (-1)^{\left(5-n_{5}\right)n_{4} + \left(5-n_{5}\right)n_{1} + \left( 1- n_{1} \right) n_{4} } *_{5} *_{1} *_{4}
	\end{displaymath}
where $n_{5}$, $n_{1}$ and $n_{4}$ are the number of indices in each part. In particular:
	\begin{eqnarray}
			 *_{10}(1) &=& \frac{1}{2} e^{\Phi} \vol{\RNAdS[5]} \wedge \left(d\psi + \mathcal{A}_{(1)} \right) \wedge J_{(2)} \wedge J_{(2)} \nonumber \\
    		 *_{5} (1) &=& \vol{\RNAdS[5]} \nonumber \\
			 *_{1} (1) &=& e^{\Phi} \left(d\psi + \mathcal{A}_{(1)} - \frac{2}{\sqrt{3}} A_{Q} \right) \nonumber \\
			 *_{4} (1) &=& \frac{1}{2} J_{(2)} \wedge J_{(2)} \nonumber \\
		*_{4}(J_{(2)}) &=& J_{(2)} \nonumber
	\end{eqnarray}
where $\vol{S^{5}} = *_{1} (1) \wedge *_{4} (1)$ when $\Phi=0$.}

{\ After Melvinisation the Hodge dual on the asymptotically \Schrcomma charged black brane spacetime, denoted \cSch[5], is not equal to that on $\RNAdS[5]$ and we must determine its effect upon our volume forms and gauge fields. It can be shown that objects whose terms all contain $dx^{-}$ pick up a factor of $e^{\Phi}$ when acted on by the Melvinised Hodge dual. After Melvinisation the following important objects Hodge dualise in the manner shown:
	\begin{eqnarray}
		\vol{\RNAdS[5]} &=& e^{-\Phi} \vol{\cSch[5]} \nonumber \\
		    *_{5} F_{Q} &=& -2 e^{-\Phi} *''_{5} \left( \frac{1}{\sqrt{3}} F_{Q}  + F \wedge A_{M} \right) \nonumber
	\end{eqnarray}
The transformation of $*_{5} F_{Q}$ was determined by considering the fact that $B''_{(2)} \wedge F''_{(3)}$ is precisely the quantity that needs to be added to make $F_{(5)}$ self-dual with respect to the Melvinised metric, see \cite{Adams:2009dm}.}

\subsection{TsT transformation}

{\ First note that generically we can write the 10-dimensional metric and five-form as:
	\begin{eqnarray}
		ds_{10}^{2} &=& g_{--} \left(dx^{-}\right)^2 + 2g_{-\alpha}dx^{-} dx^{\alpha} + g_{\alpha \beta} dx^{\alpha} dx^{\beta} + \left(d\psi + \mathcal{A}_{(1)} + A_{Q} \right)^{2} + d\Sigma_{4}^2 \nonumber \\
		F_{(5)}     &=& d\psi \wedge \left( dx^{-} \wedge A_{(3)} + B_{(4)} \right) + dx^{-} \wedge C_{(4)} + D_{(5)} \nonumber
		\label{Eq: AppendixTsT Genericforms}
	\end{eqnarray}
where $\alpha$, $\beta$ belong to $\left\{r,+,x,z\right\}$. As the TsT only ever performs algebraic operations on the $\psi$ and $x^{-}$ isometry directions we only need to keep track of these terms.}

{\ We shall need to T-dualise our solution twice so it makes sense to define a standard form for the relevant fields (as in \cite{Adams:2009dm}). In particular, we isolate all the dependence on $\psi$ in our fields and write them in the following manner:
	\begin{eqnarray}
		ds_{10}^{2} &=& g_{\psi \psi} \left(d\psi + g_{(\psi)} \right)^2 + \ldots \nonumber \\
		B_{(2)}     &=& \left(d\psi + \frac{1}{2} g_{(\psi)} \right) \wedge B_{(\psi)} + \ldots \nonumber \\
		F_{(p)}     &=& \left(d\psi + g_{(\psi)} \right) \wedge F_{(p)\psi} + F_{(p) \not\psi} \nonumber
	\end{eqnarray}
where $\not\psi$ indicates the piece of the field with no $\psi$ components. We also choose to denote the dilaton by $\Phi_{0}$. The T-dualisation of these objects then yields:
	\begin{eqnarray}
		\left(ds_{10}^{2}\right)' &=& \frac{1}{g_{\psi \psi}} \left(d\psi - B_{(\psi)} \right)^2 + \ldots \nonumber \\
					 B'_{(2)}     &=& \left(d\psi - \frac{1}{2} B_{(\psi)} \right) \wedge \left(-g_{(\psi)} \right) + \ldots \nonumber \\
		          	 F'_{(p)}     &=& \left(d\psi - B_{(\psi)} \right) \wedge \left(F_{(p-1) \not \psi}\right) +  F_{(p+1)\psi} \nonumber \\
        			 e^{\Phi'}    &=& \frac{e^{\Phi_{0}}}{g_{\psi \psi}} \nonumber
	\end{eqnarray}
We shall denote T-dualised quantities with a $'$.}

{\ As was stated in the main body of the paper the TsT transformation is formed from the following sequence of operations:
	\begin{enumerate}
		\item T-dualise along the $\psi$ direction,
		\item twist along $x^{-}$ sending it to $x^{-}+\alpha \psi$ where $\alpha$ is a constant,
		\item and finally T-dualise along the $\psi$ direction.
	\end{enumerate}
Applying these operations to the fields of (\ref{Eq: AppendixTsT Genericforms}) we obtain:
	\begin{eqnarray}
		\left(ds_{10}^{2}\right)'' &=& ds_{5}^2 + \frac{ \left(d\psi +\mathcal{A}_{(1)} - \frac{2}{\sqrt{3}} A_{Q}\right)^{2} }{k} + d\Sigma_{4}^{2} - \frac{\alpha^2}{k} \left(g_{--} dx^{-} + g_{-\alpha} dx^{\alpha} \right) 
									   \nonumber \\
					B''_{(2)}	   &=& \frac{\alpha}{k} \left(g_{--} dx^{-} + g_{-\alpha} dx^{\alpha} \right) \wedge \left(d\psi + \mathcal{A}_{(1)}  - \frac{2}{\sqrt{3}}  A_{Q} \right)  \nonumber \\
					 F''_{(3)}     &=& \alpha A_{(3)} \nonumber \\
					 F''_{(5)}     &=& F_{(5)} + B''_{(2)} \wedge F''_{(3)} \nonumber \\
					 F''_{(7)}     &=& B''_{(2)} \wedge F''_{(5)} \nonumber \\
					 e^{2\Phi''}   &=& \frac{e^{2\Phi_{0}}}{k} \nonumber
	\end{eqnarray}
where $k=1+\alpha^2 g_{--}$ and $ds^{2}_{5}$ is the original five dimensional metric. From the above formulae we can readily identify $A_{M}$ to be $\frac{\alpha}{k}g_{-\mu} dx^{\mu}$.

\bibliographystyle{JHEP}
\bibliography{references}

\end{document}